\documentclass[a4paper,aps,prb,reprint,superscriptaddress,showpacs]{revtex4-1}
\usepackage{amsmath, bm, braket}
\usepackage{graphicx}
\usepackage{hyperref}
\begin{document}
\title{Incommensurate spiral magnetic order on anisotropic triangular lattice: Dynamical mean field study in a spin-rotating frame}
\begin{abstract}
  We study the ground-state magnetism of the half-filled Hubbard model on the anisotropic triangular lattice, where two out of three bonds have hopping $t$ and the third one has $t^\prime$ in a unit triangle. Working in a spin-rotating frame and using the density matrix renormalization group method as an impurity solver, we provide a proper description  of incommensurate magnetizations at zero temperature in the framework of the dynamical mean-field theory (DMFT). It is shown that the incommensurate spiral magnetic order for $t^\prime/t\gtrsim 0.7$ survives the dynamical fluctuations of itinerant electrons in the Hubbard interaction range from the strong-coupling (localized-spin) limit down to the insulator-to-metal transition. We also find that the magnetic moment reduction from the localized-spin limit is pronounced in the vicinity of the transition between the commensurate N\'eel and incommensurate spiral phases at $t^\prime/t\sim 0.7$. When the anisotropy parameter $t^\prime/t$ increases from the N\'eel-to-spiral transition, the magnitude of the magnetic moment immediately reaches a maximum and then rapidly decreases in the range of larger $t^\prime/t$ including the isotropic triangular lattice point $t^\prime/t=1$. This work gives a solid foundation for further extension of the study including nonlocal correlation effects neglected at the standard DMFT level.  
\end{abstract}
\author{Shimpei Goto}
\email{goto@kh.phys.waseda.ac.jp}
\affiliation{Department of Physics, Waseda University, Shinjuku, Tokyo 169-8555, Japan}
\author{Susumu Kurihara}
\affiliation{Department of Physics, Waseda University, Shinjuku, Tokyo 169-8555, Japan}
\author{Daisuke Yamamoto}
\affiliation{Department of Physics and Mathematics, Aoyama-Gakuin University, Sagamihara, Kanagawa 252-5258, Japan}
\maketitle
\section{introduction}
The interplay of geometrical frustration and quantum fluctuations of itinerant electrons has drawn much attention because of its essential role for the realization of spin liquid (SL) states in organic compounds such as $\kappa$-$(\mathrm{BEDT\mathchar`-TTF})_2\mathrm{Cu}_2(\mathrm{CN})_3$, $\mathrm{EtMe}_3\mathrm{Sb[Pd(dmit)_2]_2}$, and $\kappa\mathchar`- \mathrm{H}_3(\mathrm{Cat\mathchar`-EDT\mathchar`-TTF})_2$.\cite{shimizu_spin_2003,itou_quantum_2008,isono_gapless_2014}
In these compounds, dimerized molecules form layered anisotropic triangular lattices spaced by insulating nonmagnetic layers. 
Many theoretical efforts aimed at understanding the quantum magnetism of anisotropic triangular-lattice systems have been made with the Heisenberg model of localized spins in both semi-analytical \cite{starykh_ordering_2007,hayashi_possibility_2007,herfurth_majorana_2013,merino_heisenberg_1999,trumper_spin-wave_1999,hauke_modified_2011,manuel_magnetic_1999,merino_spin-liquid_2014,ghamari_order_2011,holt_spin-liquid_2014}
and numerical\cite{bishop_magnetic_2009,harada_numerical_2012,yunoki_two_2006,heidarian_spin-$frac12$_2009,ghorbani_variational_2016,weng_spin-liquid_2006,reuther_functional_2011,thesberg_exact_2014,weichselbaum_incommensurate_2011,weihong_phase_1999} manners. 
These studies have shown that the spatial anisotropy in spin exchange interactions gives rise to an incommensurate spiral magnetic order with an irrational ordering wave vector. It has been also found that strong quantum fluctuations are induced in the anisotropy parameter range where the competition between the commensurate N\'{e}el and incommensurate spiral orders takes place or where the low dimensionality is enhanced by large anisotropy. These strong fluctuation effects could lead to quantum nonmagnetic states including SLs, although different approaches have given different conclusions\cite{merino_heisenberg_1999,trumper_spin-wave_1999,manuel_magnetic_1999,yunoki_two_2006,heidarian_spin-$frac12$_2009,ghorbani_variational_2016,weng_spin-liquid_2006,hayashi_possibility_2007,hauke_modified_2011,reuther_functional_2011,ghamari_order_2011,harada_numerical_2012,herfurth_majorana_2013,holt_spin-liquid_2014,merino_spin-liquid_2014,bishop_magnetic_2009,thesberg_exact_2014} about the anisotropy parameter range where the SL states appear.

The Hubbard model describes additional fluctuation effects that come from the itinerancy of electrons, which may also play an important role on the magnetism of the organic compounds and other strongly correlated electron systems.  However, the theoretical studies on the anisotropic triangular Hubbard model 
\cite{krishnamurthy_mott-hubbard_1990,morita_nonmagnetic_2002,dayal_absence_2012,watanabe_superconductivity_2006,tocchio_spin-liquid_2013,tocchio_one-dimensional_2014,koretsune_exact_2007,clay_absence_2008,kyung_mott_2006,ohashi_finite_2008,sahebsara_antiferromagnetism_2006,laubach_phase_2015,yamada_magnetic_2014,watanabe_predominant_2008,liebsch_mott_2009}
remain far from consensus due to the difficulty in dealing with itinerant electron systems with frustration. In order to reach full understanding of the itinerant frustrated magnetism on the anisotropic triangular lattice, it is crucial to properly treat the strong fluctuation effects between itinerant electrons and the incommensurability of magnetic orders\cite{weihong_phase_1999,powell_symmetry_2007,bishop_magnetic_2009,merino_heisenberg_1999,trumper_spin-wave_1999,weichselbaum_incommensurate_2011,merino_spin-liquid_2014,hauke_modified_2011,manuel_magnetic_1999,krishnamurthy_mott-hubbard_1990,ghamari_order_2011,holt_spin-liquid_2014,harada_numerical_2012,yunoki_two_2006,heidarian_spin-$frac12$_2009,ghorbani_variational_2016,weng_spin-liquid_2006,reuther_functional_2011,thesberg_exact_2014}. Furthermore, the consistency with the known results for the Heisenberg model of localized spins has to be achieved in the large Hubbard-interaction limit.

In this paper, we study the magnetic properties, including the incommensurability of magnetic orders, of the half-filled Hubbard model on the anisotropic triangular lattice by means of the dynamical mean field theory (DMFT) \cite{metzner_correlated_1989, georges_hubbard_1992, georges_dynamical_1996}. The DMFT treats local correlation effects between electrons in a non-perturbative fashion by mapping the original many-body problem onto an effective impurity model, which becomes exact in the limit of lattices with an infinite coordination. Therefore, the spirit of the approximation is similar to those of the Weiss molecular field theory for localized spins\cite{weiss_hypothese_1907} and the Gutzwiller approximation for lattice bosons\cite{rokhsar_gutzwiller_1991,krauth_gutzwiller_1992}. These ``single-site'' approximations have offered a good starting point for understanding the role of fluctuations in quantum many-body systems.  Based on the single-site approximations, the neglected nonlocal correlations  can be taken into account by, e.g., their cluster extensions
\cite{luhmann_cluster_2013,yamamoto_dipolar_2012,yamamoto_quantum_2014,oguchi_theory_1955,hettler_nonlocal_1998,lichtenstein_antiferromagnetism_2000,kotliar_cellular_2001} and perturbative expansions with collective-mode excitations (such as the spin-wave theory\cite{anderson_approximate_1952}).

Although several cluster extensions of the DMFT and the related approaches
\cite{ohashi_finite_2008,kyung_mott_2006,sahebsara_antiferromagnetism_2006,yamada_magnetic_2014,laubach_phase_2015,watanabe_predominant_2008,liebsch_mott_2009}
have been already applied to the Hubbard model on the anisotropic triangular lattice, the incommensurate magnetic order with irrational ordering wave vectors has not been properly treated in those studies. Here, we describe fully incommensurate orders by applying a local gauge transformation on the spin space of the electron operators. Dealing with an effective impurity model in the spin-rotating frame by means of a solver based on the density matrix renormalization group (DMRG) \cite{white_density_1992, white_density-matrix_1993, schollwock_density-matrix_2011}, we study the effects of dynamical fluctuations on the incommensurate spin spiral states in the framework of the DMFT. The zero-temperature phase diagram determined by our DMFT shows that the incommensurate magnetic order in insulating states survives the dynamical fluctuations of electrons in the interaction range from the strong-coupling (localized-spin) limit down to the insulator-to-metal transition. This indicates that it is crucial for the study of anisotropic triangular lattice to properly treat the incommensurability of the magnetic order. The role of the local, dynamic fluctuations in realizing quantum SL states in strongly correlated electron systems~\cite{shimizu_spin_2003,itou_quantum_2008} will be also discussed.

This paper is organized as follows. In Sec. \ref{sec:implementation}, we introduce the Hamiltonian of the model considered here and provide the procedure of the DMFT calculations in the spin-rotating frame. 
In Sec. \ref{sec:results}, we present the phase diagram of the model and show the behaviors of the magnetic moment and the ordering wave vector as a function of the system parameters. The role of the dynamical fluctuations in realizing the SL state is also discussed.
Conclusions are given in Sec. \ref{sec:conclusion}.
\section{dynamical mean field theory for incommensurate spiral orders}
\label{sec:implementation}
\subsection{Model Hamiltonian and the strong-coupling limit}

\begin{figure}
  \includegraphics[scale=0.4]{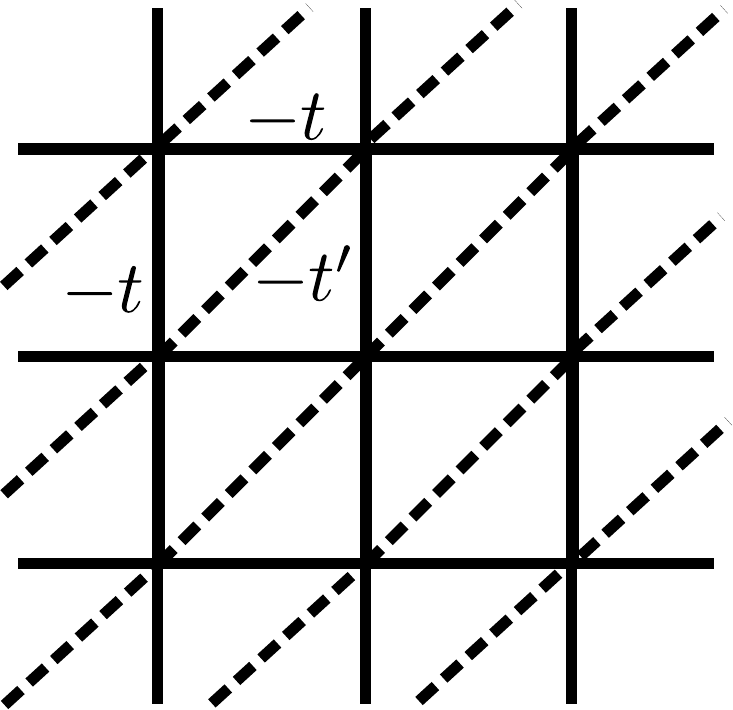}
  \caption{Square-lattice geometry that is topologically equivalent to the triangular lattice with spatially anisotropic hoppings $-t$ (solid bonds) and $-t^\prime$ (dashed bonds).}
  \label{fig:lattice}
\end{figure}

We study the half-filled Hubbard model on a spatially anisotropic triangular lattice: 
\begin{equation}
  H = \sum_{ij \sigma}t_{ij}c^\dagger_{i \sigma} c_{j \sigma} + U \sum_{i}n_{i \uparrow}n_{i \downarrow} - \mu \sum_{\sigma} n_\sigma
  \label{eq:hamiltonian},
\end{equation}
where $c_{i \sigma}$ is an annihilation operator of an electron at site $i$ with spin $\sigma$, $U$ is the on-site Hubbard interaction,
$\mu$ is the chemical potential, and $n_{i\sigma} = c^\dagger_{i\sigma}c_{i\sigma}$. The spatially anisotropic triangular lattice is equivalent to the square lattice with one additional set of diagonal bonds (see Fig.~\ref{fig:lattice}). We assume the hopping integral $t_{ij}$ as

\begin{equation}
  t_{ij}=\left\{ \begin{array}{ll}
-t <0& ({\bm r}_{j}-{\bm r}_{i}=\pm {\bm e}_1,\pm {\bm e}_2 ) \\
-t^\prime \leq 0 & ({\bm r}_{j}-{\bm r}_{i}=\pm ({\bm e}_1+{\bm e}_2)) \\
0 & ({\rm otherwise})
\end{array} \right.
\end{equation}
with ${\bm e}_1=(1,0)$, ${\bm e}_2=(0,1)$, and $\bm{r}_i$ being a position vector of site $i$. The geometry of the lattice can be viewed as an interpolation between the square lattice and the one-dimensional chain by varying $t^\prime/t$ from 0 to $\infty$ through the isotropic triangular lattice at $t^\prime/t=1$.

In the strong coupling limit of $U\gg t,t^\prime$ at half-filling, the charge degrees of freedom are frozen out, and the Hubbard model is mapped onto the Heisenberg model with exchange couplings $J=4 t^2/U$ and $J^\prime=4 (t^\prime)^2/U$ for solid and dashed bonds in Fig.~\ref{fig:lattice}, respectively. The classical-spin analysis on the anisotropic triangular Heisenberg model has shown that the local spins form a magnetic order with the ordering vector ${\bm Q}=(q,q)$ where~\cite{merino_heisenberg_1999,trumper_spin-wave_1999}
\begin{equation}
q=\left\{ \begin{array}{ll}
\arccos (-J/2J^\prime)& (J^\prime/J>1/2) \\
\pi & (J^\prime/J\leq 1/2).\label{eq:orderingvectors}
\end{array} \right.
\end{equation}
Increasing the value of $J^\prime/J$ from 0 leads to a commensurate-incommensurate transition occurs 
at $J^\prime/J=1/2$ ($t^\prime/t=1/\sqrt{2}\approx0.707$) from the N\'eel to incommensurate spiral state. 
When $J^\prime/J$ is increased further, the ordering vector takes $(2\pi/3,2\pi/3)$, 
which corresponds to a commensurate 120$^\circ$ order, 
at $J^\prime/J=1$ and approaches $(\pi/2,\pi/2)$ in the one-dimensional limit of $J^\prime/J\rightarrow \infty$.

\subsection{Dynamical mean field theory}
\label{DMFT}
Let us now turn to the discussions away from the strong coupling limit to consider the effects of charge degrees of freedom on the magnetic orders. In order to deal with the N\'{e}el and spin spiral orders within the framework of DMFT, 
we rotate the local phase of the electron operators as 
\begin{equation}
\tilde{c}_{i \sigma} = c_{i \sigma}\mathrm{e}^{\frac{\mathrm{i}\sigma}{2}(\bm{Q}\cdot \bm{r}_i + \phi)}, 
\end{equation}
where $\phi$ is an arbitrary phase shift.
Under this local gauge transformation, the Hamiltonian becomes
\begin{equation}
  \tilde{H}_{\bm{Q}} = \sum_{ij\sigma} t_{ij} \mathrm{e}^{\frac{\mathrm{i}\sigma}{2}[\bm{Q}\cdot(\bm{r}_i - \bm{r}_j)]} \tilde{c}^\dagger_{i\sigma}\tilde{c}_{j\sigma} + U\sum_{i \sigma} n_{i \uparrow}n_{i \downarrow},
\end{equation}
where $n_{i\sigma} = c^\dagger_{i\sigma}c_{i\sigma} = \tilde{c}^\dagger_{i\sigma}\tilde{c}_{i\sigma}$.
Each component of spin operator is transformed as
\begin{align}
  S^x_i &= \frac{1}{2}(c^\dagger_{i\uparrow}c_{i\downarrow} + c^\dagger_{i\downarrow}c_{i\uparrow}) \nonumber \\
  &= \frac{1}{2}(\tilde{c}^\dagger_{i\uparrow}\tilde{c}_{i\downarrow}\mathrm{e}^{\mathrm{i}(\bm{Q}\cdot \bm{r}_i + \phi)} + \tilde{c}^\dagger_{i\downarrow}\tilde{c}_{i\uparrow}\mathrm{e}^{-\mathrm{i}(\bm{Q}\cdot \bm{r}_i + \phi)})\nonumber\\ 
  &\equiv \tilde{S}^x_i \cos(\bm{Q}\cdot\bm{r}_i + \phi) - \tilde{S}^y_i \sin(\bm{Q}\cdot\bm{r}_i + \phi), \\
  S^y_i &= -\frac{\mathrm{i}}{2}(c^\dagger_{i\uparrow}c_{i\downarrow} - c^\dagger_{i\downarrow}c_{i\uparrow}) \nonumber \\
&= -\frac{\mathrm{i}}{2}(\tilde{c}^\dagger_{i\uparrow}\tilde{c}_{i\downarrow}\mathrm{e}^{\mathrm{i}(\bm{Q}\cdot \bm{r}_i + \phi)} - \tilde{c}^\dagger_{i\downarrow}\tilde{c}_{i\uparrow}\mathrm{e}^{-\mathrm{i}(\bm{Q}\cdot \bm{r}_i + \phi)})\nonumber\\ 
  &\equiv \tilde{S}^x_i \sin(\bm{Q}\cdot\bm{r}_i + \phi) + \tilde{S}^y_i \cos(\bm{Q}\cdot\bm{r}_i + \phi), \\
  S^z_i &= \frac{1}{2}(c^\dagger_{i\uparrow}c_{i\uparrow} - c^\dagger_{i\downarrow}c_{i\downarrow}) = \frac{1}{2}(\tilde{c}^\dagger_{i\uparrow}\tilde{c}_{i\uparrow} - \tilde{c}^\dagger_{i\downarrow}\tilde{c}_{i\downarrow}) \equiv \tilde{S}^z_i.
\end{align}
Therefore a magnetically ordered spiral state in the $xy$ plain can be described by a uniform magnetization, 
$\braket{\tilde{S}^x_i} = M$ and $\braket{\tilde{S}^y_i} = \braket{\tilde{S}^z_i} =0$, in the spin-rotating frame, 
which is convenient for the DMFT formulation of the system with incommensurate spiral orders. 

It is expected for finite $U$ that due to the charge fluctuation effects, 
the magnetization $M$ is reduced and the ordering vector $\bm{Q}=(q_x, q_y)$ is shifted from the classical-spin result in Eq.~(\ref{eq:orderingvectors}). The local Green's function for $\tilde{c}_{i\sigma}$ is given by
\begin{eqnarray}
\bm{G}(\omega)&=&\left( \begin{array}{cc}
\langle\langle\tilde{c}_{i\uparrow};\tilde{c}^\dagger_{i\uparrow}\rangle\rangle_\omega& \langle\langle\tilde{c}_{i\uparrow};\tilde{c}^\dagger_{i\downarrow}\rangle\rangle_\omega \\
\langle\langle\tilde{c}_{i\downarrow};\tilde{c}^\dagger_{i\uparrow}\rangle\rangle_\omega & \langle\langle\tilde{c}_{i\downarrow};\tilde{c}^\dagger_{i\downarrow}\rangle\rangle_\omega
\end{array} \right)\\\nonumber
&=&\frac{1}{N}\sum_{\bm{k}}\frac{1}{(\omega+\mu)\openone-
\bm{\varepsilon}_{\bm{Q}}(\bm{k})
-\bm{\Sigma}({\bm{k},\omega})},\label{Green's function}
\end{eqnarray}
where $N$ is the number of lattice sites 
and $\bm{\varepsilon}_{\bm{Q}}(\bm{k})$ is a diagonal matrix whose component 
$\displaystyle \varepsilon_{\bm{Q}\sigma \sigma}(\bm{k})= -2t \sum_{\nu=x,y}\cos\left(k_\nu + \sigma \frac{q_\nu}{2}\right) - 2t^\prime \cos\left(\sum_{\nu=x,y}\left(k_\nu+\sigma \frac{q_\nu}{2}\right)\right)$ 
is the single-particle dispersion of $\tilde{c}_{i\sigma}$. 
The effects of spatial and dynamical fluctuations induced by the interactions $U$ are taken into account 
through the momentum $\bm{k}=(k_x, k_y)$ and frequency $\omega$ dependences of the self-energy $\bm{\Sigma}(\bm{k}, \omega)$. 
In the simple DMFT, the self-energy is approximated as $\bm{\Sigma}(\bm{k}, \omega) \approx \bm{\Sigma}(\omega)$ to study the local fluctuation effects. 
Under the approximation, the problem is mapped onto the single impurity Anderson model (SIAM),\cite{georges_dynamical_1996} whose Hamiltonian is given by

\begin{align}
  H_{\mathrm{SIAM}} &= U n_{\uparrow} n_{\downarrow} - \mu \sum_{\sigma}n_{\sigma} 
  \nonumber \\* &+  
  \sum_{l\sigma \sigma^\prime}^{N_b} (V_{l\sigma \sigma^\prime}a^\dagger_{l \sigma} \tilde{c}_{\sigma^\prime} + \mathrm{H.c.}) + \sum_{l\sigma}^{N_b} \epsilon_l a^\dagger_{l \sigma} a_{l \sigma},~\label{eq:SIAM}
\end{align}
where $\tilde{c}_{\sigma}$ is an annihilation operator of an electron at impurity site with spin $\sigma$, $n_{\sigma} = \tilde{c}^\dagger_{\sigma} \tilde{c}_{\sigma}$, 
$a_{l\sigma}$ is an annihilation operator of an electron at $l$-th bath orbital with spin $\sigma$, and $N_b$ is the number of bath orbitals. The bath parameters $V_{l\sigma \sigma^\prime}$ and $\varepsilon_l$ should be optimized so that the impurity Green's function 
\begin{eqnarray}
  \bm{G}_{\rm imp}(\omega)=\frac{1}{(\omega+\mu)\openone-\bm{\Gamma}(\omega)-\bm{\Sigma}({\omega})},\label{eq:G_imp}
\end{eqnarray}
is equal to the local Green's function $\bm{G}(\omega)$ of the original lattice problem [Eq.~(\ref{Green's function})] with the replacement of $\bm{\Sigma}(\bm{k}, \omega)$ by $\bm{\Sigma}(\omega)$. Here, the hybridization function $\bm{\Gamma}(\omega)$ is given by
\begin{equation}
  \bm{\Gamma}(\omega) = \sum_l \frac{\bm{V}_l\bm{V}^\dagger_l}{\omega - \epsilon_l},\label{eq:hyb}
\end{equation}
where $\bm{V}_l$ is a two-by-two matrix whose component is $V_{l\sigma \sigma^\prime}$. The spin-flip couplings $V_{l \uparrow\downarrow}$ and $V_{l \downarrow\uparrow}$ are required to describe the in-plane magnetization $M=\langle\tilde{S^x}\rangle$.

In order to compute the impurity Green's function {$\bm{G}_\mathrm{imp}(\omega)$}, 
we employ the imaginary-time matrix product state solver \cite{wolf_imaginary-time_2015} based on the DMRG technique, which can treat dozens of bath orbitals and access zero temperature. In the DMRG calculations, which provide the ground state of the system, the SIAM Hamiltonian is arranged in the star geometry,\cite{wolf_solving_2014} and the truncation error is set to lower than $10^{-8}$. The imaginary-time Green's function $\bm{G}_\mathrm{imp}(\tau)$ can be computed from a one-electron (one-hole) excited state,\cite{wolf_imaginary-time_2015} which is obtained by applying a creation (annihilation) operator to the ground state. For an efficient Fourier transformation of the Green's function with respect to $\tau$, we perform the fitting of each component of $\bm{G}_\mathrm{imp}(\tau)$ in the form $\sum_i \alpha_i \mathrm{e}^{-\beta_i \tau}$ with the matrix pencil method. \cite{sarkar_using_1995} This procedure gives the impurity Green's function $\bm{G}_\mathrm{imp}(\omega) $ on the imaginary axis for a given set of the bath parameters $V_{l\sigma \sigma^\prime}$ and $\varepsilon_l$. The details of the optimization of the bath parameters under the condition $\bm{G}_\mathrm{imp}(\omega)=\bm{G}(\omega)$ are given in the Appendix.

In addition to the self-consistent optimization of the bath parameters, one has to determine spin spiral ordering vector $\bm{Q}$ so that the energy of the system can be minimized with respect to $\bm{Q}$. The energy of the system $E(\bm{Q})$ as a function of $\bm{Q}$ is given by the Galitskii-Migdal formula,\cite{fetter_quantum_2003}
\begin{equation}
  E(\bm{Q}) = \frac{1}{N}\sum_{\bm{k}}\int_{C} \frac{\mathrm{d}\omega}{2\pi \mathrm{i}} \mathrm{Tr}
  \left[ 
    \left(
      \bm{\varepsilon}_{\bm{Q}}(\bm{k}) + \frac{1}{2}\bm{\Sigma}(\omega)
    \right)
    \bm{G}_{\mathrm{latt}}(\bm{k}, \omega)
  \right]. \label{eq:migdal}
\end{equation}
Here, $C$ denotes a contour which surrounds the negative real axis counterclockwise 
and $\bm{G}_{\mathrm{latt}}(\bm{k}, \omega)$ is the lattice Green's function of the DMFT which is given by
\begin{equation}
  \bm{G}_{\mathrm{latt}}(\bm{k}, \omega) = \frac{1}{(\omega+\mu)\openone - \bm{\varepsilon}_{\bm{Q}}(\bm{k})-\bm{\Sigma}(\omega)}.
\end{equation}
This contour integration can be transformed into an integration over the positive imaginary axis.\cite{senechal_introduction_2008} 
Note that the minimization of the energy function $E(\bm{Q})$ with respect to $\bm{Q}$ can be also obtained by the stability condition
\begin{equation}
  \frac{\partial}{\partial q_{\nu}} \braket{\tilde{H}_{\bm{Q}}} = \braket{{j}_{\nu \bm{Q}}} = 0~~(\nu=x,y),
\end{equation}
where ${j}_{\nu \bm{Q}}\equiv \frac{\partial \tilde{H}_{\bm{Q}}}{\partial q_{\nu}} = 
\sum_{ij \sigma} \mathrm{i}\sigma \frac{q_\nu}{2} t_{ij}(\nu_i - \nu_j)\tilde{c}^\dagger_{i\sigma}\tilde{c}_{j\sigma}$ 
is the spin current operator in the $\nu$ direction. 
Here, $\nu_i$ is the $\nu$-component of the vector $\bm{r}_i = (x_i, y_i)$.

The local quantities including the filling $\sum_\sigma \langle {n}_{\sigma}\rangle$ and the spin moments $\langle \tilde{\bm{S}}\rangle$ can be directly calculated from the local Green's function $\bm{G}(\omega)$ with the optimized values of the bath parameters and the ordering vector $\bm{Q}$. In order to consider the half-filled case, the chemical potential $\mu$ has to be numerically tuned so that $\sum_\sigma \langle {n}_{\sigma}\rangle=1$ since the system for $t,t^\prime\neq 0$ does not possess the particle-hole symmetry.

Using the above-mentioned DMFT procedure in the spin-rotating frame, one can describe the insulating state with an incommensurate spiral magnetic order and the commensurate N\'{e}el and 120$^\circ$ antiferromagnetic states, as well as metallic states. In the followings, we will mainly discuss the charge fluctuation effects on the magnetic properties of the insulating states in the region of large but finite values of $U$. The possibility of the $d$-wave superconducting state~\cite{kyung_mott_2006} for intermediate $U/t$ is out of the scope of this paper since spatial correlations are neglected. 

\section{magnetic orders and metal-insulator transitions}
\label{sec:results}
\subsection{Magnetic phase diagram}
\begin{figure}[b]
  \includegraphics[scale=1.0]{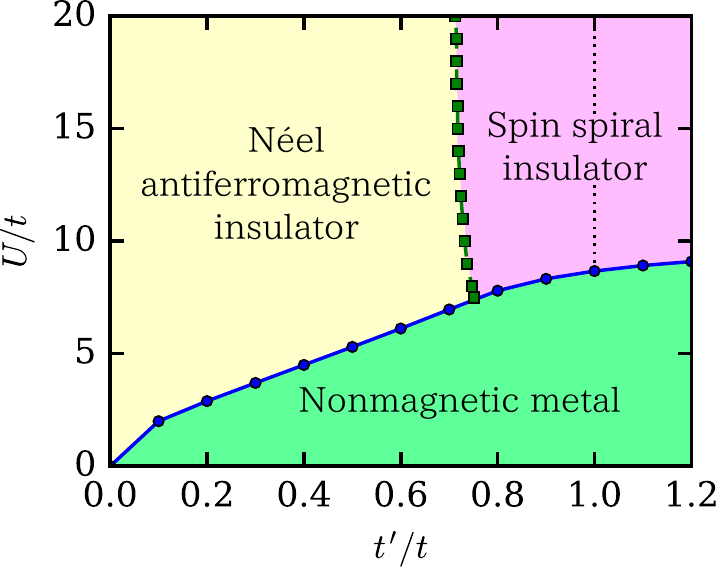}
  \caption{(Color online) Magnetic phase diagram of the half-filled Hubbard model on the anisotropic triangular lattice.
  The line with blue circles (green squares) represents a first-order (second-order) transition boundary. The spin spiral phase has an incommensurate magnetic order except at $t^\prime/t=1$ (dashed line)}
  \label{fig:phase_diag}
\end{figure}
\begin{figure}
  \includegraphics[scale=0.75]{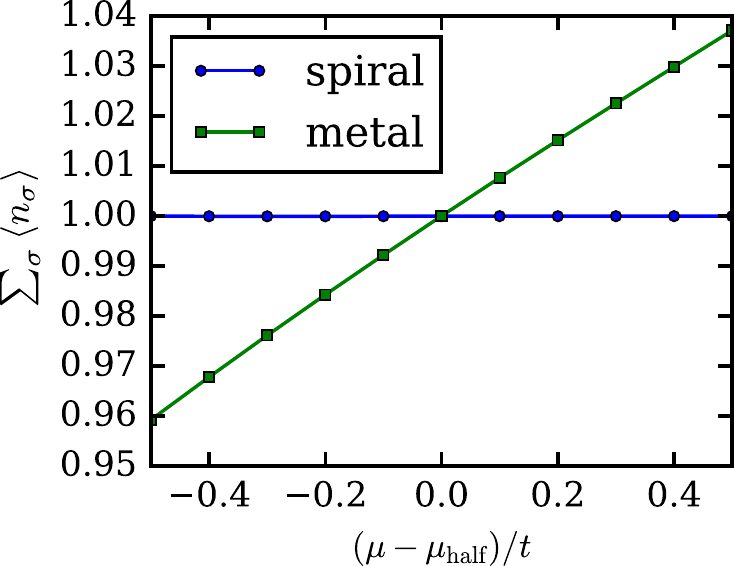}
  \caption{(Color online) Chemical potential $\mu$ dependences of filling $\sum_\sigma \braket{n_\sigma}$ at $t^\prime/t = 0.9$.
    The line with blue circles (green squares) corresponds to spin spiral (metal) phase at $U/t=9.0$ ($U/t=8.0$).
  Here, $\mu_\mathrm{half}$ is the value of the chemical potential when $\sum_\sigma \braket{n_\sigma}=1$}.
  \label{fig:fill}
\end{figure}

In Fig.~\ref{fig:phase_diag} we show the ground-state phase diagram obtained by the DMFT calculations in the spin-rotating frame. 
The phase diagram consists of three phases: the N\'{e}el-antiferromagnetic and spin-spiral insulators as well as a nonmagnetic-metal phase. The magnetic orders of the former two are characterized by the ordering vector $\bm{Q} =(\pi,\pi)$ and $\bm{Q} =(q, q)$ with $\pi/2<q<\pi$, respectively. In Fig.~\ref{fig:fill}, we show the chemical potential dependence of the filling $\sum_\sigma \braket{n_\sigma}$ for a typical spin-spiral insulator and metallic states. It can be seen that the slope is zero in a finite range of $\mu$ in the spin spiral state, which indicates the opening of a charge gap. 

\begin{figure}
  \includegraphics[scale=0.75]{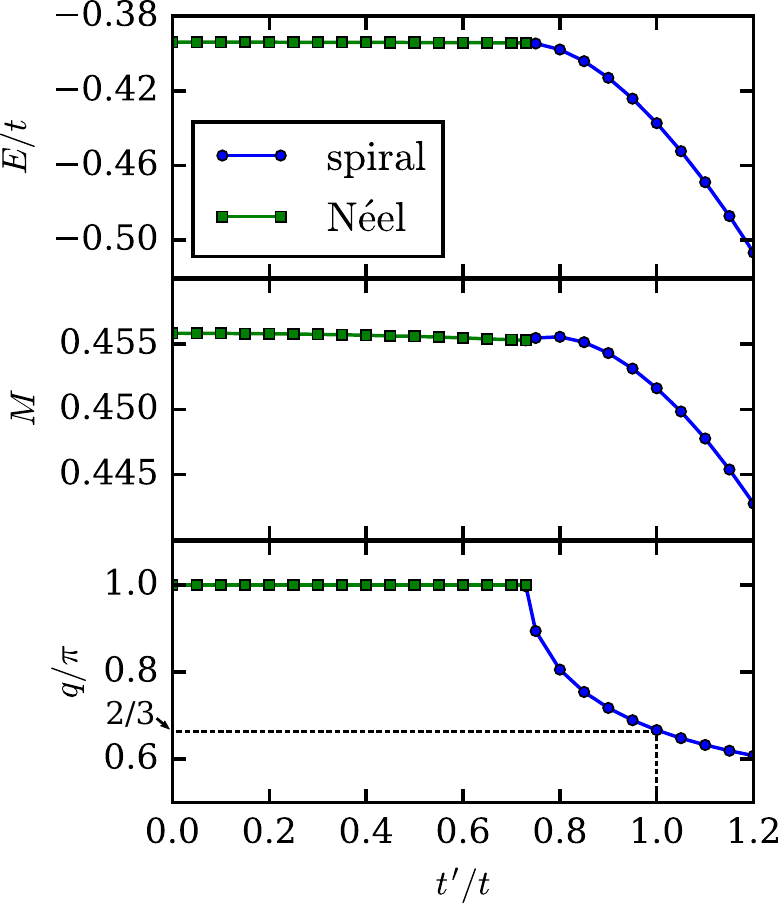}
  \caption{(Color online) Upper panel: diagonal hopping $t^\prime$ dependence of energy per site at $U/t = 10$. 
    Middle panel: the hopping dependence of magnetic moment $M$.
    Lower panel: the hopping dependence of ordering vector parameter $q$.}
  \label{fig:neel-spiral}
\end{figure}
\begin{figure}
  \includegraphics[scale=0.75]{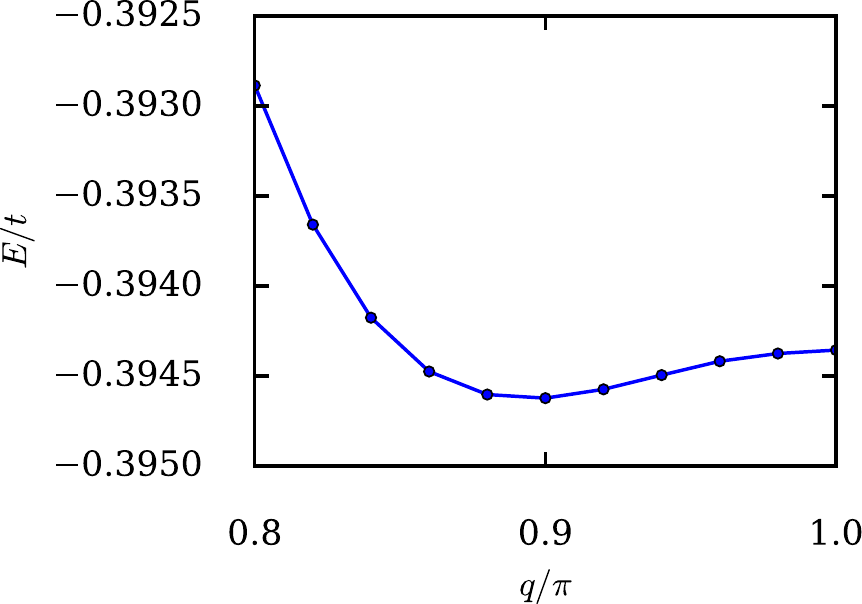}
  \caption{(Color online) The ordering vector dependence of the energy function $E(q,q)$ at $U/t=10.0$ and $t^\prime/t = 0.75$.
  The bath parameters are optimized for each value of $q$.}
  \label{fig:q_ene}
\end{figure}
Figure~\ref{fig:neel-spiral} shows how the anisotropy $t^\prime/t$ affects the magnetic orders in the insulator phases at strong interactions. When $t^\prime/t=0$, the system is reduced to the simple square-lattice Hubbard model, which is well-known to exhibit a robust N\'{e}el order due to the perfect nesting of the itinerant electron Fermi surface. As shown in the lower panel of Fig.~\ref{fig:neel-spiral}, even if the lattice geometry is changed by finite $t^\prime/t$, the N\'{e}el order with commensurate wave vector $(\pi,\pi)$ persists up to a certain critical value $(t^\prime/t)_c$. For $t^\prime/t>(t^\prime/t)_c$, the minimum of the energy function $E(\bm{Q})$ is shifted from $(\pi,\pi)$ to an incommensurate momentum $(q,q)$ as shown in Fig.~\ref{fig:q_ene}, which indicates a transition to a state with an incommensurate magnetic order. As $t^\prime/t$ increases, the value of $q$ continuously moves away from $\pi$ and reaches $2\pi/3$ at the isotropic triangular-lattice point $t^\prime/t=1$. The wave vector $\bm{Q} =(2\pi/3,2\pi/3)$ corresponds to a commensurate (three-sublattice) 120$^\circ$ order expected in triangular-lattice antiferromagnetic systems.\cite{bernu_exact_1994,capriotti_long-range_1999,white_neorder_2007,shirakawa_ground_2016}
For a dominant diagonal hopping $t^\prime>t$, the value of $q$ further decreases and approaches $\pi/2$ in the one-dimensional limit of $t^\prime>t\rightarrow \infty$.

This behavior of magnetic order as a function of the anisotropy $t^\prime/t$ for large $U/t$ is consistent with the classical-spin analysis of the antiferromagnetic Heisenberg model on the anisotropic triangular lattice.~\cite{merino_heisenberg_1999, trumper_spin-wave_1999, merino_spin-liquid_2014} In fact, the ordering vector $(q,q)$ and the magnetic moment $M$ approach the classical-spin results, Eq.~(\ref{eq:orderingvectors}) with $J^\prime/J=(t^\prime/t)^2$ and $M=S=1/2$, in the limit of the infinite Hubbard interaction $U/t\rightarrow\infty$. This agreement is not surprising since the DMFT neglects the spatial fluctuations (the $\bm{k}$ dependence) in the self energy $\bm{\Sigma}(\bm{k}, \omega)$ as in classical-spin systems. Therefore, the reduction of the magnetic moment $M$ shown in the middle panel of Fig.~\ref{fig:neel-spiral} is purely the result of the local, dynamical fluctuations that stem from the itinerant charge degrees of freedom. The magnetic moment $M$ exhibits a dip at the transition point between the commensurate N\'{e}el and incommensurate spiral phases, although the reduction from $M=S$ is at most only $\sim 10$ percent. In the spiral phase, the curve of $M$ shows a peak (at $t^\prime/t\sim 0.8$ in the case of Fig.~\ref{fig:neel-spiral}), and then decreases as $t^\prime/t$ increases.

\begin{figure}
  \includegraphics[scale=0.75]{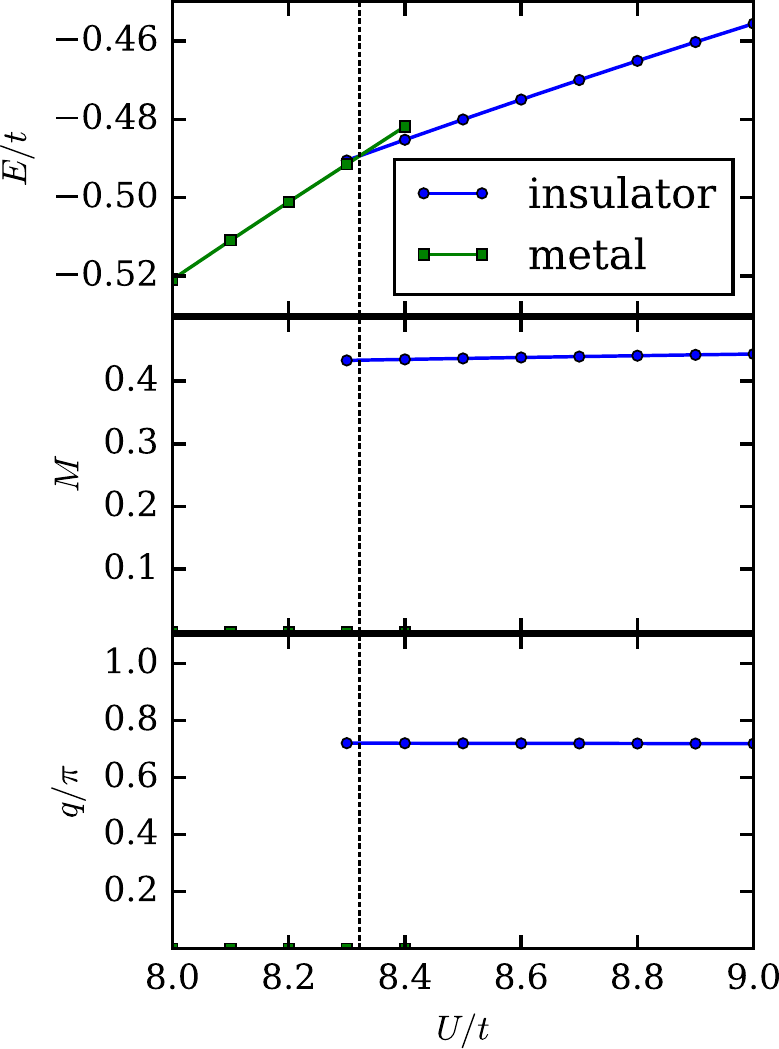}
  \caption{(Color online) Upper panel: Hubbard interaction $U$ dependence of energy per site at $t^\prime/t = 0.9$. 
    The vertical dashed line represents the first-order transition point. Middle panel: the interaction dependence of magnetic moment $M$.
    Lower panel: the interaction dependence of ordering vector parameter $q$.}
  \label{fig:Spiral}
\end{figure}
On decreasing the interaction $U/t$, the system with $t^\prime/t>0$ undergoes a first-order transition from a magnetic insulator to a metallic state at a certain value of $U/t$. This is because the perfect nesting condition of the half-filled square lattice is violated for $t^\prime/t\neq 0$, and finite $U/t$ is required to stabilize magnetic orders. As shown in Fig.~\ref{fig:Spiral}, the magnetic moment $M$ suddenly vanishes at the metal-insulator transition point. In our DMFT analysis, no magnetic metal state is found between the magnetic insulator and nonmagnetic metal phases in the parameter range of the phase diagram in Fig.~\ref{fig:phase_diag}. This is consistent with the previous studies in Refs.~[\onlinecite{sahebsara_antiferromagnetism_2006,clay_absence_2008,dayal_absence_2012,kyung_mott_2006,morita_nonmagnetic_2002,watanabe_superconductivity_2006,koretsune_exact_2007,ohashi_finite_2008,watanabe_predominant_2008,sahebsara_hubbard_2008,liebsch_mott_2009,tocchio_spin-liquid_2013,tocchio_one-dimensional_2014,laubach_phase_2015}], although several works including the Hartree-Fock mean-field analysis~\cite{krishnamurthy_mott-hubbard_1990} and the variational cluster approach\cite{yamada_magnetic_2014} have predicted the existence of magnetic metal phases for intermediate interactions.

\subsection{Possible spin liquid: Spatial and dynamical fluctuations}

The possibility of quantum SLs on anisotropic triangular lattice has been discussed in both localized-spin systems~
\cite{merino_heisenberg_1999,trumper_spin-wave_1999,manuel_magnetic_1999,yunoki_two_2006,heidarian_spin-$frac12$_2009,ghorbani_variational_2016,weng_spin-liquid_2006,hayashi_possibility_2007,hauke_modified_2011,reuther_functional_2011,ghamari_order_2011,harada_numerical_2012,herfurth_majorana_2013,holt_spin-liquid_2014,merino_spin-liquid_2014,bishop_magnetic_2009,thesberg_exact_2014} 
and insulating yet barely itinerant electrons.~\cite{morita_nonmagnetic_2002,clay_absence_2008,dayal_absence_2012,sahebsara_antiferromagnetism_2006,kyung_mott_2006,watanabe_superconductivity_2006,koretsune_exact_2007,ohashi_finite_2008,watanabe_predominant_2008,sahebsara_hubbard_2008,liebsch_mott_2009,tocchio_spin-liquid_2013,tocchio_one-dimensional_2014,yamada_magnetic_2014,laubach_phase_2015} 
In those strongly correlated electron systems, two types of quantum fluctuation effects play a key role for ``quantum melting'' of conventional magnetic long-range order: strong spatial fluctuations due to the frustrated lattice geometry and dynamical charge and spin fluctuations due to the itinerancy of electrons.

The former effects have been studied in terms of the Heisenberg model of localized spins with anisotropic exchange $J$ and $J^\prime$ (or the half-filled Hubbard model in the large $U/t$ limit).\cite{weihong_phase_1999,starykh_ordering_2007,merino_heisenberg_1999,trumper_spin-wave_1999,manuel_magnetic_1999,yunoki_two_2006,weng_spin-liquid_2006,hayashi_possibility_2007,heidarian_spin-$frac12$_2009,hauke_modified_2011,reuther_functional_2011,herfurth_majorana_2013,holt_spin-liquid_2014,thesberg_exact_2014,ghorbani_variational_2016,ghamari_order_2011,harada_numerical_2012,merino_spin-liquid_2014,bishop_magnetic_2009} 
Of particular interest is the anisotropy range where the classical spin configuration changes from the commensurate N\'{e}el to incommensurate spiral phase. 
The linear spin-wave theory has shown that the spin-wave velocity along the $(k,k)$ direction vanishes at the N\'{e}el-spiral transition,\cite{merino_heisenberg_1999,trumper_spin-wave_1999} 
which indicates that the magnetic order is destroyed by long-wavelength excitations. 
However, different approximations including several types of spin-wave theories,\cite{hauke_modified_2011} 
Schwinger-boson mean-field method,\cite{merino_spin-liquid_2014} and series-expansion approach~\cite{weihong_phase_1999} have led to different conclusions on the search of SL phases in this anisotropy region, and more sophisticated numerical studies~
\cite{bishop_magnetic_2009,ghorbani_variational_2016} have been very limited. In the region where the anisotropic triangular lattice can be regarded as weakly-coupled chains ($J^\prime/J> 1$), the fate of the classical spiral state under the influence of quantum fluctuations has been examined by various numerical calculations, which have suggested the emergence of nontrivial ground states including essentially one-dimensional (gapless) SLs,\cite{yunoki_two_2006,hayashi_possibility_2007,heidarian_spin-$frac12$_2009,weng_spin-liquid_2006,reuther_functional_2011,herfurth_majorana_2013,ghorbani_variational_2016} 
a gapped SL close to the isotropic point, \cite{yunoki_two_2006,heidarian_spin-$frac12$_2009} 
and a collinear antiferromagnetically ordered state.\cite{starykh_ordering_2007,ghamari_order_2011}

On the other hand, the effects of the local, dynamical fluctuations unique to itinerant electrons has been discussed separately from the spatial fluctuations in our present DMFT analysis on the Hubbard model with finite values of $U/t$. 
For the Hubbard model, the previous study with a cellular DMFT~\cite{kyung_mott_2006} has shown that a nonmagnetic SL state may appear in a wide range of the anisotropy parameter, $0.9\lesssim t^\prime /t<1.2$, for large $U/t$. 
However, it should be noted that such cluster-based approximations~\cite{sahebsara_antiferromagnetism_2006,kyung_mott_2006,yamada_magnetic_2014,laubach_phase_2015} can describe only a commensurate magnetic order allowed by the size of the assumed cluster (four sites in Ref.~[\onlinecite{kyung_mott_2006}]). 
The phase diagram obtained by our DMFT in the spin-rotating frame (Fig.~\ref{fig:phase_diag}) shows that the incommensurate spiral phase persists until it undergoes a first-order transition to the metallic phase, and no SL phase is formed only by the local quantum fluctuations due to the itinerant change degrees of freedom.

\begin{figure}[b]
  \includegraphics[scale=0.75]{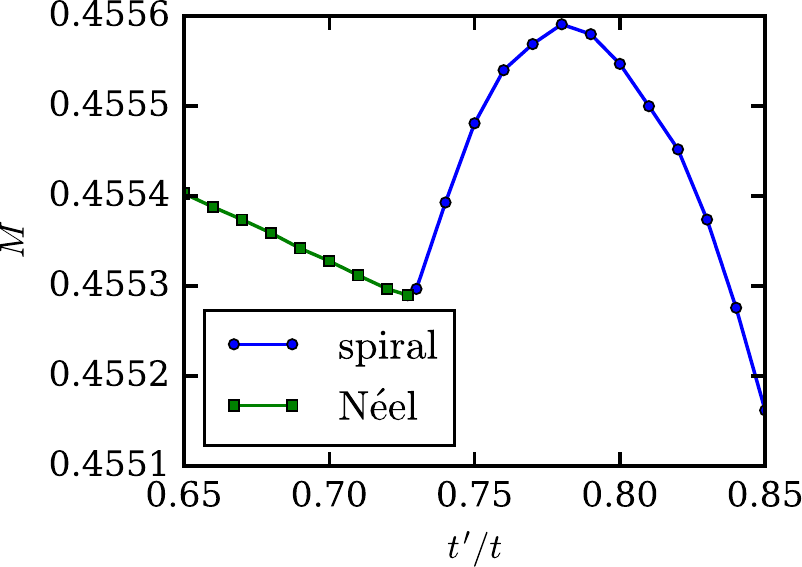}
  \caption{(Color online) The hopping $t^\prime$ dependence of the magnetic moment $M$ around the N\'{e}el-spiral transition point.}
  \label{fig:enlarged}
\end{figure}
In order to reach the final conclusion on the ground-state magnetic property of the Hubbard model for generic values of $t^\prime/t$, it is required to take into consideration the interplay of both the spatial and dynamical fluctuations and compare the energies of incommensurate spiral state and SL (or the other candidate) states. Our DMFT calculations provide valuable insight to solve this problem. Figure~\ref{fig:enlarged} is the enlarged view of the middle panel of Fig.~\ref{fig:neel-spiral}. The figure shows that the reduction of the magnetic moment due to the local fluctuations of itinerant electrons is pronounced in the vicinity of the N\'{e}el-spiral transition. Therefore, the itinerant charge degrees of freedom should work in the direction to help the emergence of SL expected in the same region of the Heisenberg model.

Another interesting feature is observed regarding to the robustness of the spiral magnetic order for larger values of $t^\prime/t$. The linear spin-wave analysis,\cite{merino_heisenberg_1999,trumper_spin-wave_1999} the Schwinger boson mean-field approach,\cite{merino_spin-liquid_2014} and the coupled cluster method~\cite{bishop_magnetic_2009} for the Heisenberg model have all shown that the spatial fluctuations on the spiral order due to the frustrated geometry is most suppressed at the isotropic triangular-lattice point, and the 120$^\circ$ magnetic order is robust against the fluctuations. On the other hand, the curve of $M$ in Fig.~\ref{fig:enlarged} exhibits a maximum at a small value of $t^\prime/t$ apart from the isotropic triangular-lattice point and is rapidly decreasing for larger values of $t^\prime/t$. This result indicates that the fluctuations coming from the itinerancy of electrons have significant effects in the anisotropy range including the isotropic triangular-lattice point as well as in the essentially one-dimensional region of $t^\prime/t\gg 1$. This fact may support the scenario that the finite $U/t$ effects could induce a SL ground state even at the isotropic triangular-lattice point.\cite{morita_nonmagnetic_2002,kyung_mott_2006,dayal_absence_2012,clay_absence_2008,sahebsara_antiferromagnetism_2006,watanabe_superconductivity_2006,koretsune_exact_2007,ohashi_finite_2008,yamada_magnetic_2014,laubach_phase_2015} 
\section{conclusion}
\label{sec:conclusion}
In this paper, we studied the effects of the itinerant electron degrees of freedom on the magnetic properties of the systems on the anisotropic triangular lattice that interpolates from the square lattice ($t^\prime/t= 0$) to decoupled one-dimensional chains ($t^\prime/t\rightarrow \infty$) via the isotropic triangular lattice ($t^\prime/t=1$). We performed a local gauge transformation that rotated the spin-quantization axis into the direction of the magnetic moment at each site to properly describe an incommensurate spin spiral order. Working in the spin-rotating frame and using the imaginary-time matrix product state solver~\cite{wolf_imaginary-time_2015} based on the DMRG, we determined the magnetic phase diagram of the half-filled anisotropic-triangular Hubbard model at zero temperature in the framework of the DMFT. It was found that the metal-insulator transition for $t^\prime/t\neq 0$ takes place at a nonzero value of $U/t$ due to the lack of perfect nesting, and in a discontinuous (first-order) fashion. When the anisotropy parameter $t^\prime/t$ increases from $0$ in the insulating state at a fixed value of $U/t$, the ordering vector of the magnetic long-range order changes from the rational value $(\pi,\pi)$ to an irrational one $(q,q)$ at $t^\prime/t\sim 0.7$, and gradually goes to $(\pi/2,\pi/2)$ as $t^\prime/t\rightarrow \infty$.

In the vicinity of the transition between the commensurate N\'{e}el and incommensurate spiral states, the magnetic moment reduction caused by the fluctuation effects is pronounced. Moreover, for large values of $t^\prime/t$, the magnetic moment decreases rapidly with $t^\prime/t$ due to the enhancement of low dimensionality. It is noteworthy that such a strong reduction of the magnetic moment already begins at the isotropic triangular-lattice point $t^\prime/t=1$ unlike the case of the quantum spin fluctuations of localized spin systems.\cite{merino_heisenberg_1999,trumper_spin-wave_1999,bishop_magnetic_2009,merino_spin-liquid_2014} As shown in the phase diagram of Fig.~\ref{fig:phase_diag}, no nonmagnetic insulating state was formed only by the local, dynamic electron fluctuations considered in the DMFT. This indicates that spatial quantum fluctuations are required for the emergence of SL states. Nevertheless, our calculations predict that the itinerant electron fluctuations for finite $U/t$ could help the emergence of SL states in the vicinity of the commensurate-incommensurate transition and in a large-$t^\prime/t$ region, which might include the isotropic triangular-lattice point.\cite{morita_nonmagnetic_2002,kyung_mott_2006,dayal_absence_2012,clay_absence_2008,sahebsara_antiferromagnetism_2006,watanabe_superconductivity_2006,koretsune_exact_2007,ohashi_finite_2008,yamada_magnetic_2014,laubach_phase_2015}

The inclusion of nonlocal fluctuation effects has been partially carried out by cluster extensions of DMFT,\cite{kyung_mott_2006,ohashi_finite_2008} which have, however, treated only commensurate magnetic orders allowed within the assumed cluster shape. As was pointed out in the present study, incommensurability of magnetic order is essential for the magnetic property of the anisotropic triangular-lattice systems, and moreover, long-wavelength fluctuations are important for the breaking of long-range magnetic orders according to linear spin-wave predictions.\cite{merino_heisenberg_1999,trumper_spin-wave_1999} Our present DMFT calculations in the spin-rotating frame provide a solid physical and mathematical basis for further study in this direction, e.g., with diagrammatic extensions of DMFT,\cite{toschi_dynamical_2007, rubtsov_dual_2008} which can include the effects of long-range quantum correlations through diagrammatic correction.

\begin{acknowledgments}
  The DMRG calculations in this paper are performed using ITensor library, http://itensor.org.
  This paper is a part of the outcome of research performed under a Waseda University Grant for Special Research Projects No. 2015S-100 (S.G.) and partially supported by KAKENHI Grants from Japan Society for the Promotion of Science No. 26800200 (D.Y.).
\end{acknowledgments}
\appendix*
\section{How to optimize bath parameters}
From Eqs.~\eqref{eq:G_imp} and \eqref{eq:hyb}, the self-consistent condition of the DMFT, $\bm{G}_\mathrm{imp}(\omega) = \bm{G}(\omega)$, can be rewritten as
\begin{equation}
  \sum_l \frac{\bm{V}_l\bm{V}^\dagger_l}{\omega - \epsilon_l} = 
  \omega + \mu - \bm{\Sigma}(\omega) - \bm{G}^{-1}(\omega). \label{eq:condition}
\end{equation}
Using Eq.~(\ref{eq:condition}) we adjust the bath parameters $V_{l\sigma \sigma^\prime}$ and $\varepsilon_l$ in an iterative manner: First, the SIAM in Eq.~(\ref{eq:SIAM}) is solved by the DMRG technique given in Sec.~\ref{DMFT}, and the self-energy $\bm{\Sigma}(\omega)$ is extracted by the calculated $\bm{G}_\mathrm{imp}(\omega)$ via Eq.~(\ref{eq:G_imp}). Substituting $\bm{\Sigma}(\omega)$, one can evaluate the right-hand side of Eq.~(\ref{eq:condition}). (Note that the self-energy $\bm{\Sigma}({\bm k},\omega)$ in $\bm{G}(\omega)$ should be replaced by $\bm{\Sigma}(\omega)$ in the DMFT.) Then a new set of $V_{l\sigma \sigma^\prime}$ and $\varepsilon_l$ is given by fitting the evaluated right-hand-side value in the form of the left-hand side as a function of $\omega$. Using the updated bath parameters we solve again the SIAM by the DMRG technique, and the procedure is repeated until convergence is reached. The convergence criterion used in this study is 
\begin{equation}
\sum_\omega \left \| \bm{\Gamma}(\omega) - \bm{\Gamma}^\prime(\omega)\right \| < 5 \times 10^{-3}t,
\end{equation} 
where $\bm{\Gamma}(\omega)$ and $\bm{\Gamma}^\prime(\omega)$ are the hybridization function $\sum_l (\bm{V}_l\bm{V}^\dagger_l)/(\omega - \epsilon_l)$ with the bath parameters before and after a single step of the DMFT iteration. Here we take the summation over a set of 200 sample points $\omega=(0.1\mathrm{i}t,0.2\mathrm{i}t,\cdots,20\mathrm{i}t)$ on the imaginary axis.

The fitting of both sides of Eq.~(\ref{eq:condition}) for updating the bath parameters is performed by minimizing the distance function
\begin{equation}
  d = \sum_\omega \left \| \sum_l \frac{\bm{V}_{l}\bm{V}^\dagger_{l}}{\omega - \epsilon_l} - 
  \left[\omega + \mu - \bm{\Sigma}(\omega) - \bm{G}^{-1}(\omega)\right]\right \| . \label{eq:distance}
\end{equation}
Since the distance function is nonconvex, the minimization by ordinary gradient methods is practically difficult (See the supplemental material of Ref. [\onlinecite{go_spatial_2015}]).
Thus, to perform the minimization in an efficient way, we use the vector fitting (VF) method\cite{gustavsen_rational_1999,gustavsen_modal_2008}, which gives a fitting of the numerical data for the right-hand-side of Eq.~(\ref{eq:condition}) with a rational expression $\sum_l \bm{A}_l/(\omega - \epsilon_l)$. The matrix $\bm{V}_l$ can be obtained by the Cholesky decomposition of the matrix $\bm{A}_l$. It should be noted that if the number of the bath orbitals $N_b$ (the number of the bath parameters) is too large, the VF method may provide a non-positive definite matrix $\bm{A}_l$, which cannot be decomposed by the Cholesky decomposition, and/or a complex value for $\epsilon_l$ due to ``overfitting.'' To avoid it, we try the fittings with different $N_b$ (typically up to $N_b \sim 25$ in the present study), and choose the best fitting out of them. The value of the distance function $d$ for the 200 $\omega$ points is smaller than $10^{-4}t$ throughout the calculations.

\bibliography{my_library}

\begin{thebibliography}{71}%
\makeatletter
\providecommand \@ifxundefined [1]{%
 \@ifx{#1\undefined}
}%
\providecommand \@ifnum [1]{%
 \ifnum #1\expandafter \@firstoftwo
 \else \expandafter \@secondoftwo
 \fi
}%
\providecommand \@ifx [1]{%
 \ifx #1\expandafter \@firstoftwo
 \else \expandafter \@secondoftwo
 \fi
}%
\providecommand \natexlab [1]{#1}%
\providecommand \enquote  [1]{``#1''}%
\providecommand \bibnamefont  [1]{#1}%
\providecommand \bibfnamefont [1]{#1}%
\providecommand \citenamefont [1]{#1}%
\providecommand \href@noop [0]{\@secondoftwo}%
\providecommand \href [0]{\begingroup \@sanitize@url \@href}%
\providecommand \@href[1]{\@@startlink{#1}\@@href}%
\providecommand \@@href[1]{\endgroup#1\@@endlink}%
\providecommand \@sanitize@url [0]{\catcode `\\12\catcode `\$12\catcode
  `\&12\catcode `\#12\catcode `\^12\catcode `\_12\catcode `\%12\relax}%
\providecommand \@@startlink[1]{}%
\providecommand \@@endlink[0]{}%
\providecommand \url  [0]{\begingroup\@sanitize@url \@url }%
\providecommand \@url [1]{\endgroup\@href {#1}{\urlprefix }}%
\providecommand \urlprefix  [0]{URL }%
\providecommand \Eprint [0]{\href }%
\providecommand \doibase [0]{http://dx.doi.org/}%
\providecommand \selectlanguage [0]{\@gobble}%
\providecommand \bibinfo  [0]{\@secondoftwo}%
\providecommand \bibfield  [0]{\@secondoftwo}%
\providecommand \translation [1]{[#1]}%
\providecommand \BibitemOpen [0]{}%
\providecommand \bibitemStop [0]{}%
\providecommand \bibitemNoStop [0]{.\EOS\space}%
\providecommand \EOS [0]{\spacefactor3000\relax}%
\providecommand \BibitemShut  [1]{\csname bibitem#1\endcsname}%
\let\auto@bib@innerbib\@empty
\bibitem [{\citenamefont {Shimizu}\ \emph {et~al.}(2003)\citenamefont
  {Shimizu}, \citenamefont {Miyagawa}, \citenamefont {Kanoda}, \citenamefont
  {Maesato},\ and\ \citenamefont {Saito}}]{shimizu_spin_2003}%
  \BibitemOpen
  \bibfield  {author} {\bibinfo {author} {\bibfnamefont {Y.}~\bibnamefont
  {Shimizu}}, \bibinfo {author} {\bibfnamefont {K.}~\bibnamefont {Miyagawa}},
  \bibinfo {author} {\bibfnamefont {K.}~\bibnamefont {Kanoda}}, \bibinfo
  {author} {\bibfnamefont {M.}~\bibnamefont {Maesato}}, \ and\ \bibinfo
  {author} {\bibfnamefont {G.}~\bibnamefont {Saito}},\ }\href {\doibase
  10.1103/PhysRevLett.91.107001} {\bibfield  {journal} {\bibinfo  {journal}
  {Phys. Rev. Lett.}\ }\textbf {\bibinfo {volume} {91}},\ \bibinfo {pages}
  {107001} (\bibinfo {year} {2003})}\BibitemShut {NoStop}%
\bibitem [{\citenamefont {Itou}\ \emph {et~al.}(2008)\citenamefont {Itou},
  \citenamefont {Oyamada}, \citenamefont {Maegawa}, \citenamefont {Tamura},\
  and\ \citenamefont {Kato}}]{itou_quantum_2008}%
  \BibitemOpen
  \bibfield  {author} {\bibinfo {author} {\bibfnamefont {T.}~\bibnamefont
  {Itou}}, \bibinfo {author} {\bibfnamefont {A.}~\bibnamefont {Oyamada}},
  \bibinfo {author} {\bibfnamefont {S.}~\bibnamefont {Maegawa}}, \bibinfo
  {author} {\bibfnamefont {M.}~\bibnamefont {Tamura}}, \ and\ \bibinfo {author}
  {\bibfnamefont {R.}~\bibnamefont {Kato}},\ }\href {\doibase
  10.1103/PhysRevB.77.104413} {\bibfield  {journal} {\bibinfo  {journal} {Phys.
  Rev. B}\ }\textbf {\bibinfo {volume} {77}},\ \bibinfo {pages} {104413}
  (\bibinfo {year} {2008})}\BibitemShut {NoStop}%
\bibitem [{\citenamefont {Isono}\ \emph {et~al.}(2014)\citenamefont {Isono},
  \citenamefont {Kamo}, \citenamefont {Ueda}, \citenamefont {Takahashi},
  \citenamefont {Kimata}, \citenamefont {Tajima}, \citenamefont {Tsuchiya},
  \citenamefont {Terashima}, \citenamefont {Uji},\ and\ \citenamefont
  {Mori}}]{isono_gapless_2014}%
  \BibitemOpen
  \bibfield  {author} {\bibinfo {author} {\bibfnamefont {T.}~\bibnamefont
  {Isono}}, \bibinfo {author} {\bibfnamefont {H.}~\bibnamefont {Kamo}},
  \bibinfo {author} {\bibfnamefont {A.}~\bibnamefont {Ueda}}, \bibinfo {author}
  {\bibfnamefont {K.}~\bibnamefont {Takahashi}}, \bibinfo {author}
  {\bibfnamefont {M.}~\bibnamefont {Kimata}}, \bibinfo {author} {\bibfnamefont
  {H.}~\bibnamefont {Tajima}}, \bibinfo {author} {\bibfnamefont
  {S.}~\bibnamefont {Tsuchiya}}, \bibinfo {author} {\bibfnamefont
  {T.}~\bibnamefont {Terashima}}, \bibinfo {author} {\bibfnamefont
  {S.}~\bibnamefont {Uji}}, \ and\ \bibinfo {author} {\bibfnamefont
  {H.}~\bibnamefont {Mori}},\ }\href {\doibase 10.1103/PhysRevLett.112.177201}
  {\bibfield  {journal} {\bibinfo  {journal} {Phys. Rev. Lett.}\ }\textbf
  {\bibinfo {volume} {112}},\ \bibinfo {pages} {177201} (\bibinfo {year}
  {2014})}\BibitemShut {NoStop}%
\bibitem [{\citenamefont {Starykh}\ and\ \citenamefont
  {Balents}(2007)}]{starykh_ordering_2007}%
  \BibitemOpen
  \bibfield  {author} {\bibinfo {author} {\bibfnamefont {O.~A.}\ \bibnamefont
  {Starykh}}\ and\ \bibinfo {author} {\bibfnamefont {L.}~\bibnamefont
  {Balents}},\ }\href {\doibase 10.1103/PhysRevLett.98.077205} {\bibfield
  {journal} {\bibinfo  {journal} {Phys. Rev. Lett.}\ }\textbf {\bibinfo
  {volume} {98}},\ \bibinfo {pages} {077205} (\bibinfo {year}
  {2007})}\BibitemShut {NoStop}%
\bibitem [{\citenamefont {Hayashi}\ and\ \citenamefont
  {Ogata}(2007)}]{hayashi_possibility_2007}%
  \BibitemOpen
  \bibfield  {author} {\bibinfo {author} {\bibfnamefont {Y.}~\bibnamefont
  {Hayashi}}\ and\ \bibinfo {author} {\bibfnamefont {M.}~\bibnamefont
  {Ogata}},\ }\href {\doibase 10.1143/JPSJ.76.053705} {\bibfield  {journal}
  {\bibinfo  {journal} {J. Phys. Soc. Jpn.}\ }\textbf {\bibinfo {volume}
  {76}},\ \bibinfo {pages} {053705} (\bibinfo {year} {2007})}\BibitemShut
  {NoStop}%
\bibitem [{\citenamefont {Herfurth}\ \emph {et~al.}(2013)\citenamefont
  {Herfurth}, \citenamefont {Streib},\ and\ \citenamefont
  {Kopietz}}]{herfurth_majorana_2013}%
  \BibitemOpen
  \bibfield  {author} {\bibinfo {author} {\bibfnamefont {T.}~\bibnamefont
  {Herfurth}}, \bibinfo {author} {\bibfnamefont {S.}~\bibnamefont {Streib}}, \
  and\ \bibinfo {author} {\bibfnamefont {P.}~\bibnamefont {Kopietz}},\ }\href
  {\doibase 10.1103/PhysRevB.88.174404} {\bibfield  {journal} {\bibinfo
  {journal} {Phys. Rev. B}\ }\textbf {\bibinfo {volume} {88}},\ \bibinfo
  {pages} {174404} (\bibinfo {year} {2013})}\BibitemShut {NoStop}%
\bibitem [{\citenamefont {Merino}\ \emph {et~al.}(1999)\citenamefont {Merino},
  \citenamefont {McKenzie}, \citenamefont {Marston},\ and\ \citenamefont
  {Chung}}]{merino_heisenberg_1999}%
  \BibitemOpen
  \bibfield  {author} {\bibinfo {author} {\bibfnamefont {J.}~\bibnamefont
  {Merino}}, \bibinfo {author} {\bibfnamefont {R.~H.}\ \bibnamefont
  {McKenzie}}, \bibinfo {author} {\bibfnamefont {J.~B.}\ \bibnamefont
  {Marston}}, \ and\ \bibinfo {author} {\bibfnamefont {C.~H.}\ \bibnamefont
  {Chung}},\ }\href {\doibase 10.1088/0953-8984/11/14/012} {\bibfield
  {journal} {\bibinfo  {journal} {J. Phys.: Condens. Matter}\ }\textbf
  {\bibinfo {volume} {11}},\ \bibinfo {pages} {2965} (\bibinfo {year}
  {1999})}\BibitemShut {NoStop}%
\bibitem [{\citenamefont {Trumper}(1999)}]{trumper_spin-wave_1999}%
  \BibitemOpen
  \bibfield  {author} {\bibinfo {author} {\bibfnamefont {A.~E.}\ \bibnamefont
  {Trumper}},\ }\href {\doibase 10.1103/PhysRevB.60.2987} {\bibfield  {journal}
  {\bibinfo  {journal} {Phys. Rev. B}\ }\textbf {\bibinfo {volume} {60}},\
  \bibinfo {pages} {2987} (\bibinfo {year} {1999})}\BibitemShut {NoStop}%
\bibitem [{\citenamefont {Hauke}\ \emph {et~al.}(2011)\citenamefont {Hauke},
  \citenamefont {Roscilde}, \citenamefont {Murg}, \citenamefont {Cirac},\ and\
  \citenamefont {Schmied}}]{hauke_modified_2011}%
  \BibitemOpen
  \bibfield  {author} {\bibinfo {author} {\bibfnamefont {P.}~\bibnamefont
  {Hauke}}, \bibinfo {author} {\bibfnamefont {T.}~\bibnamefont {Roscilde}},
  \bibinfo {author} {\bibfnamefont {V.}~\bibnamefont {Murg}}, \bibinfo {author}
  {\bibfnamefont {J.~I.}\ \bibnamefont {Cirac}}, \ and\ \bibinfo {author}
  {\bibfnamefont {R.}~\bibnamefont {Schmied}},\ }\href {\doibase
  10.1088/1367-2630/13/7/075017} {\bibfield  {journal} {\bibinfo  {journal}
  {New J. Phys.}\ }\textbf {\bibinfo {volume} {13}},\ \bibinfo {pages} {075017}
  (\bibinfo {year} {2011})}\BibitemShut {NoStop}%
\bibitem [{\citenamefont {Manuel}\ and\ \citenamefont
  {Ceccatto}(1999)}]{manuel_magnetic_1999}%
  \BibitemOpen
  \bibfield  {author} {\bibinfo {author} {\bibfnamefont {L.~O.}\ \bibnamefont
  {Manuel}}\ and\ \bibinfo {author} {\bibfnamefont {H.~A.}\ \bibnamefont
  {Ceccatto}},\ }\href {\doibase 10.1103/PhysRevB.60.9489} {\bibfield
  {journal} {\bibinfo  {journal} {Phys. Rev. B}\ }\textbf {\bibinfo {volume}
  {60}},\ \bibinfo {pages} {9489} (\bibinfo {year} {1999})}\BibitemShut
  {NoStop}%
\bibitem [{\citenamefont {Merino}\ \emph {et~al.}(2014)\citenamefont {Merino},
  \citenamefont {Holt},\ and\ \citenamefont
  {Powell}}]{merino_spin-liquid_2014}%
  \BibitemOpen
  \bibfield  {author} {\bibinfo {author} {\bibfnamefont {J.}~\bibnamefont
  {Merino}}, \bibinfo {author} {\bibfnamefont {M.}~\bibnamefont {Holt}}, \ and\
  \bibinfo {author} {\bibfnamefont {B.~J.}\ \bibnamefont {Powell}},\ }\href
  {\doibase 10.1103/PhysRevB.89.245112} {\bibfield  {journal} {\bibinfo
  {journal} {Phys. Rev. B}\ }\textbf {\bibinfo {volume} {89}},\ \bibinfo
  {pages} {245112} (\bibinfo {year} {2014})}\BibitemShut {NoStop}%
\bibitem [{\citenamefont {Ghamari}\ \emph {et~al.}(2011)\citenamefont
  {Ghamari}, \citenamefont {Kallin}, \citenamefont {Lee},\ and\ \citenamefont
  {S{\o}rensen}}]{ghamari_order_2011}%
  \BibitemOpen
  \bibfield  {author} {\bibinfo {author} {\bibfnamefont {S.}~\bibnamefont
  {Ghamari}}, \bibinfo {author} {\bibfnamefont {C.}~\bibnamefont {Kallin}},
  \bibinfo {author} {\bibfnamefont {S.-S.}\ \bibnamefont {Lee}}, \ and\
  \bibinfo {author} {\bibfnamefont {E.~S.}\ \bibnamefont {S{\o}rensen}},\
  }\href {\doibase 10.1103/PhysRevB.84.174415} {\bibfield  {journal} {\bibinfo
  {journal} {Phys. Rev. B}\ }\textbf {\bibinfo {volume} {84}},\ \bibinfo
  {pages} {174415} (\bibinfo {year} {2011})}\BibitemShut {NoStop}%
\bibitem [{\citenamefont {Holt}\ \emph {et~al.}(2014)\citenamefont {Holt},
  \citenamefont {Powell},\ and\ \citenamefont
  {Merino}}]{holt_spin-liquid_2014}%
  \BibitemOpen
  \bibfield  {author} {\bibinfo {author} {\bibfnamefont {M.}~\bibnamefont
  {Holt}}, \bibinfo {author} {\bibfnamefont {B.~J.}\ \bibnamefont {Powell}}, \
  and\ \bibinfo {author} {\bibfnamefont {J.}~\bibnamefont {Merino}},\ }\href
  {\doibase 10.1103/PhysRevB.89.174415} {\bibfield  {journal} {\bibinfo
  {journal} {Phys. Rev. B}\ }\textbf {\bibinfo {volume} {89}},\ \bibinfo
  {pages} {174415} (\bibinfo {year} {2014})}\BibitemShut {NoStop}%
\bibitem [{\citenamefont {Bishop}\ \emph {et~al.}(2009)\citenamefont {Bishop},
  \citenamefont {Li}, \citenamefont {Farnell},\ and\ \citenamefont
  {Campbell}}]{bishop_magnetic_2009}%
  \BibitemOpen
  \bibfield  {author} {\bibinfo {author} {\bibfnamefont {R.~F.}\ \bibnamefont
  {Bishop}}, \bibinfo {author} {\bibfnamefont {P.~H.~Y.}\ \bibnamefont {Li}},
  \bibinfo {author} {\bibfnamefont {D.~J.~J.}\ \bibnamefont {Farnell}}, \ and\
  \bibinfo {author} {\bibfnamefont {C.~E.}\ \bibnamefont {Campbell}},\ }\href
  {\doibase 10.1103/PhysRevB.79.174405} {\bibfield  {journal} {\bibinfo
  {journal} {Phys. Rev. B}\ }\textbf {\bibinfo {volume} {79}},\ \bibinfo
  {pages} {174405} (\bibinfo {year} {2009})}\BibitemShut {NoStop}%
\bibitem [{\citenamefont {Harada}(2012)}]{harada_numerical_2012}%
  \BibitemOpen
  \bibfield  {author} {\bibinfo {author} {\bibfnamefont {K.}~\bibnamefont
  {Harada}},\ }\href {\doibase 10.1103/PhysRevB.86.184421} {\bibfield
  {journal} {\bibinfo  {journal} {Phys. Rev. B}\ }\textbf {\bibinfo {volume}
  {86}},\ \bibinfo {pages} {184421} (\bibinfo {year} {2012})}\BibitemShut
  {NoStop}%
\bibitem [{\citenamefont {Yunoki}\ and\ \citenamefont
  {Sorella}(2006)}]{yunoki_two_2006}%
  \BibitemOpen
  \bibfield  {author} {\bibinfo {author} {\bibfnamefont {S.}~\bibnamefont
  {Yunoki}}\ and\ \bibinfo {author} {\bibfnamefont {S.}~\bibnamefont
  {Sorella}},\ }\href {\doibase 10.1103/PhysRevB.74.014408} {\bibfield
  {journal} {\bibinfo  {journal} {Phys. Rev. B}\ }\textbf {\bibinfo {volume}
  {74}},\ \bibinfo {pages} {014408} (\bibinfo {year} {2006})}\BibitemShut
  {NoStop}%
\bibitem [{\citenamefont {Heidarian}\ \emph {et~al.}(2009)\citenamefont
  {Heidarian}, \citenamefont {Sorella},\ and\ \citenamefont
  {Becca}}]{heidarian_spin-$frac12$_2009}%
  \BibitemOpen
  \bibfield  {author} {\bibinfo {author} {\bibfnamefont {D.}~\bibnamefont
  {Heidarian}}, \bibinfo {author} {\bibfnamefont {S.}~\bibnamefont {Sorella}},
  \ and\ \bibinfo {author} {\bibfnamefont {F.}~\bibnamefont {Becca}},\ }\href
  {\doibase 10.1103/PhysRevB.80.012404} {\bibfield  {journal} {\bibinfo
  {journal} {Phys. Rev. B}\ }\textbf {\bibinfo {volume} {80}},\ \bibinfo
  {pages} {012404} (\bibinfo {year} {2009})}\BibitemShut {NoStop}%
\bibitem [{\citenamefont {Ghorbani}\ \emph {et~al.}(2016)\citenamefont
  {Ghorbani}, \citenamefont {Tocchio},\ and\ \citenamefont
  {Becca}}]{ghorbani_variational_2016}%
  \BibitemOpen
  \bibfield  {author} {\bibinfo {author} {\bibfnamefont {E.}~\bibnamefont
  {Ghorbani}}, \bibinfo {author} {\bibfnamefont {L.~F.}\ \bibnamefont
  {Tocchio}}, \ and\ \bibinfo {author} {\bibfnamefont {F.}~\bibnamefont
  {Becca}},\ }\href {\doibase 10.1103/PhysRevB.93.085111} {\bibfield  {journal}
  {\bibinfo  {journal} {Phys. Rev. B}\ }\textbf {\bibinfo {volume} {93}},\
  \bibinfo {pages} {085111} (\bibinfo {year} {2016})}\BibitemShut {NoStop}%
\bibitem [{\citenamefont {Weng}\ \emph {et~al.}(2006)\citenamefont {Weng},
  \citenamefont {Sheng}, \citenamefont {Weng},\ and\ \citenamefont
  {Bursill}}]{weng_spin-liquid_2006}%
  \BibitemOpen
  \bibfield  {author} {\bibinfo {author} {\bibfnamefont {M.~Q.}\ \bibnamefont
  {Weng}}, \bibinfo {author} {\bibfnamefont {D.~N.}\ \bibnamefont {Sheng}},
  \bibinfo {author} {\bibfnamefont {Z.~Y.}\ \bibnamefont {Weng}}, \ and\
  \bibinfo {author} {\bibfnamefont {R.~J.}\ \bibnamefont {Bursill}},\ }\href
  {\doibase 10.1103/PhysRevB.74.012407} {\bibfield  {journal} {\bibinfo
  {journal} {Phys. Rev. B}\ }\textbf {\bibinfo {volume} {74}},\ \bibinfo
  {pages} {012407} (\bibinfo {year} {2006})}\BibitemShut {NoStop}%
\bibitem [{\citenamefont {Reuther}\ and\ \citenamefont
  {Thomale}(2011)}]{reuther_functional_2011}%
  \BibitemOpen
  \bibfield  {author} {\bibinfo {author} {\bibfnamefont {J.}~\bibnamefont
  {Reuther}}\ and\ \bibinfo {author} {\bibfnamefont {R.}~\bibnamefont
  {Thomale}},\ }\href {\doibase 10.1103/PhysRevB.83.024402} {\bibfield
  {journal} {\bibinfo  {journal} {Phys. Rev. B}\ }\textbf {\bibinfo {volume}
  {83}},\ \bibinfo {pages} {024402} (\bibinfo {year} {2011})}\BibitemShut
  {NoStop}%
\bibitem [{\citenamefont {Thesberg}\ and\ \citenamefont
  {S{\o}rensen}(2014)}]{thesberg_exact_2014}%
  \BibitemOpen
  \bibfield  {author} {\bibinfo {author} {\bibfnamefont {M.}~\bibnamefont
  {Thesberg}}\ and\ \bibinfo {author} {\bibfnamefont {E.~S.}\ \bibnamefont
  {S{\o}rensen}},\ }\href {\doibase 10.1103/PhysRevB.90.115117} {\bibfield
  {journal} {\bibinfo  {journal} {Phys. Rev. B}\ }\textbf {\bibinfo {volume}
  {90}},\ \bibinfo {pages} {115117} (\bibinfo {year} {2014})}\BibitemShut
  {NoStop}%
\bibitem [{\citenamefont {Weichselbaum}\ and\ \citenamefont
  {White}(2011)}]{weichselbaum_incommensurate_2011}%
  \BibitemOpen
  \bibfield  {author} {\bibinfo {author} {\bibfnamefont {A.}~\bibnamefont
  {Weichselbaum}}\ and\ \bibinfo {author} {\bibfnamefont {S.~R.}\ \bibnamefont
  {White}},\ }\href {\doibase 10.1103/PhysRevB.84.245130} {\bibfield  {journal}
  {\bibinfo  {journal} {Phys. Rev. B}\ }\textbf {\bibinfo {volume} {84}},\
  \bibinfo {pages} {245130} (\bibinfo {year} {2011})}\BibitemShut {NoStop}%
\bibitem [{\citenamefont {Weihong}\ \emph {et~al.}(1999)\citenamefont
  {Weihong}, \citenamefont {McKenzie},\ and\ \citenamefont
  {Singh}}]{weihong_phase_1999}%
  \BibitemOpen
  \bibfield  {author} {\bibinfo {author} {\bibfnamefont {Z.}~\bibnamefont
  {Weihong}}, \bibinfo {author} {\bibfnamefont {R.~H.}\ \bibnamefont
  {McKenzie}}, \ and\ \bibinfo {author} {\bibfnamefont {R.~R.~P.}\ \bibnamefont
  {Singh}},\ }\href {\doibase 10.1103/PhysRevB.59.14367} {\bibfield  {journal}
  {\bibinfo  {journal} {Phys. Rev. B}\ }\textbf {\bibinfo {volume} {59}},\
  \bibinfo {pages} {14367} (\bibinfo {year} {1999})}\BibitemShut {NoStop}%
\bibitem [{\citenamefont {Krishnamurthy}\ \emph {et~al.}(1990)\citenamefont
  {Krishnamurthy}, \citenamefont {Jayaprakash}, \citenamefont {Sarker},\ and\
  \citenamefont {Wenzel}}]{krishnamurthy_mott-hubbard_1990}%
  \BibitemOpen
  \bibfield  {author} {\bibinfo {author} {\bibfnamefont {H.~R.}\ \bibnamefont
  {Krishnamurthy}}, \bibinfo {author} {\bibfnamefont {C.}~\bibnamefont
  {Jayaprakash}}, \bibinfo {author} {\bibfnamefont {S.}~\bibnamefont {Sarker}},
  \ and\ \bibinfo {author} {\bibfnamefont {W.}~\bibnamefont {Wenzel}},\ }\href
  {\doibase 10.1103/PhysRevLett.64.950} {\bibfield  {journal} {\bibinfo
  {journal} {Phys. Rev. Lett.}\ }\textbf {\bibinfo {volume} {64}},\ \bibinfo
  {pages} {950} (\bibinfo {year} {1990})}\BibitemShut {NoStop}%
\bibitem [{\citenamefont {Morita}\ \emph {et~al.}(2002)\citenamefont {Morita},
  \citenamefont {Watanabe},\ and\ \citenamefont
  {Imada}}]{morita_nonmagnetic_2002}%
  \BibitemOpen
  \bibfield  {author} {\bibinfo {author} {\bibfnamefont {H.}~\bibnamefont
  {Morita}}, \bibinfo {author} {\bibfnamefont {S.}~\bibnamefont {Watanabe}}, \
  and\ \bibinfo {author} {\bibfnamefont {M.}~\bibnamefont {Imada}},\ }\href
  {\doibase 10.1143/JPSJ.71.2109} {\bibfield  {journal} {\bibinfo  {journal}
  {J. Phys. Soc. Jpn.}\ }\textbf {\bibinfo {volume} {71}},\ \bibinfo {pages}
  {2109} (\bibinfo {year} {2002})}\BibitemShut {NoStop}%
\bibitem [{\citenamefont {Dayal}\ \emph {et~al.}(2012)\citenamefont {Dayal},
  \citenamefont {Clay},\ and\ \citenamefont {Mazumdar}}]{dayal_absence_2012}%
  \BibitemOpen
  \bibfield  {author} {\bibinfo {author} {\bibfnamefont {S.}~\bibnamefont
  {Dayal}}, \bibinfo {author} {\bibfnamefont {R.~T.}\ \bibnamefont {Clay}}, \
  and\ \bibinfo {author} {\bibfnamefont {S.}~\bibnamefont {Mazumdar}},\ }\href
  {\doibase 10.1103/PhysRevB.85.165141} {\bibfield  {journal} {\bibinfo
  {journal} {Phys. Rev. B}\ }\textbf {\bibinfo {volume} {85}},\ \bibinfo
  {pages} {165141} (\bibinfo {year} {2012})}\BibitemShut {NoStop}%
\bibitem [{\citenamefont {Watanabe}\ \emph {et~al.}(2006)\citenamefont
  {Watanabe}, \citenamefont {Yokoyama}, \citenamefont {Tanaka},\ and\
  \citenamefont {Inoue}}]{watanabe_superconductivity_2006}%
  \BibitemOpen
  \bibfield  {author} {\bibinfo {author} {\bibfnamefont {T.}~\bibnamefont
  {Watanabe}}, \bibinfo {author} {\bibfnamefont {H.}~\bibnamefont {Yokoyama}},
  \bibinfo {author} {\bibfnamefont {Y.}~\bibnamefont {Tanaka}}, \ and\ \bibinfo
  {author} {\bibfnamefont {J.-i.}\ \bibnamefont {Inoue}},\ }\href {\doibase
  10.1143/JPSJ.75.074707} {\bibfield  {journal} {\bibinfo  {journal} {J. Phys.
  Soc. Jpn.}\ }\textbf {\bibinfo {volume} {75}},\ \bibinfo {pages} {074707}
  (\bibinfo {year} {2006})}\BibitemShut {NoStop}%
\bibitem [{\citenamefont {Tocchio}\ \emph {et~al.}(2013)\citenamefont
  {Tocchio}, \citenamefont {Feldner}, \citenamefont {Becca}, \citenamefont
  {Valent{\'i}},\ and\ \citenamefont {Gros}}]{tocchio_spin-liquid_2013}%
  \BibitemOpen
  \bibfield  {author} {\bibinfo {author} {\bibfnamefont {L.~F.}\ \bibnamefont
  {Tocchio}}, \bibinfo {author} {\bibfnamefont {H.}~\bibnamefont {Feldner}},
  \bibinfo {author} {\bibfnamefont {F.}~\bibnamefont {Becca}}, \bibinfo
  {author} {\bibfnamefont {R.}~\bibnamefont {Valent{\'i}}}, \ and\ \bibinfo
  {author} {\bibfnamefont {C.}~\bibnamefont {Gros}},\ }\href {\doibase
  10.1103/PhysRevB.87.035143} {\bibfield  {journal} {\bibinfo  {journal} {Phys.
  Rev. B}\ }\textbf {\bibinfo {volume} {87}},\ \bibinfo {pages} {035143}
  (\bibinfo {year} {2013})}\BibitemShut {NoStop}%
\bibitem [{\citenamefont {Tocchio}\ \emph {et~al.}(2014)\citenamefont
  {Tocchio}, \citenamefont {Gros}, \citenamefont {Valent{\'i}},\ and\
  \citenamefont {Becca}}]{tocchio_one-dimensional_2014}%
  \BibitemOpen
  \bibfield  {author} {\bibinfo {author} {\bibfnamefont {L.~F.}\ \bibnamefont
  {Tocchio}}, \bibinfo {author} {\bibfnamefont {C.}~\bibnamefont {Gros}},
  \bibinfo {author} {\bibfnamefont {R.}~\bibnamefont {Valent{\'i}}}, \ and\
  \bibinfo {author} {\bibfnamefont {F.}~\bibnamefont {Becca}},\ }\href
  {\doibase 10.1103/PhysRevB.89.235107} {\bibfield  {journal} {\bibinfo
  {journal} {Phys. Rev. B}\ }\textbf {\bibinfo {volume} {89}},\ \bibinfo
  {pages} {235107} (\bibinfo {year} {2014})}\BibitemShut {NoStop}%
\bibitem [{\citenamefont {Koretsune}\ \emph {et~al.}(2007)\citenamefont
  {Koretsune}, \citenamefont {Motome},\ and\ \citenamefont
  {Furusaki}}]{koretsune_exact_2007}%
  \BibitemOpen
  \bibfield  {author} {\bibinfo {author} {\bibfnamefont {T.}~\bibnamefont
  {Koretsune}}, \bibinfo {author} {\bibfnamefont {Y.}~\bibnamefont {Motome}}, \
  and\ \bibinfo {author} {\bibfnamefont {A.}~\bibnamefont {Furusaki}},\ }\href
  {\doibase 10.1143/JPSJ.76.074719} {\bibfield  {journal} {\bibinfo  {journal}
  {J. Phys. Soc. Jpn.}\ }\textbf {\bibinfo {volume} {76}},\ \bibinfo {pages}
  {074719} (\bibinfo {year} {2007})}\BibitemShut {NoStop}%
\bibitem [{\citenamefont {Clay}\ \emph {et~al.}(2008)\citenamefont {Clay},
  \citenamefont {Li},\ and\ \citenamefont {Mazumdar}}]{clay_absence_2008}%
  \BibitemOpen
  \bibfield  {author} {\bibinfo {author} {\bibfnamefont {R.~T.}\ \bibnamefont
  {Clay}}, \bibinfo {author} {\bibfnamefont {H.}~\bibnamefont {Li}}, \ and\
  \bibinfo {author} {\bibfnamefont {S.}~\bibnamefont {Mazumdar}},\ }\href
  {\doibase 10.1103/PhysRevLett.101.166403} {\bibfield  {journal} {\bibinfo
  {journal} {Phys. Rev. Lett.}\ }\textbf {\bibinfo {volume} {101}},\ \bibinfo
  {pages} {166403} (\bibinfo {year} {2008})}\BibitemShut {NoStop}%
\bibitem [{\citenamefont {Kyung}\ and\ \citenamefont
  {Tremblay}(2006)}]{kyung_mott_2006}%
  \BibitemOpen
  \bibfield  {author} {\bibinfo {author} {\bibfnamefont {B.}~\bibnamefont
  {Kyung}}\ and\ \bibinfo {author} {\bibfnamefont {A.-M.~S.}\ \bibnamefont
  {Tremblay}},\ }\href {\doibase 10.1103/PhysRevLett.97.046402} {\bibfield
  {journal} {\bibinfo  {journal} {Phys. Rev. Lett.}\ }\textbf {\bibinfo
  {volume} {97}},\ \bibinfo {pages} {046402} (\bibinfo {year}
  {2006})}\BibitemShut {NoStop}%
\bibitem [{\citenamefont {Ohashi}\ \emph {et~al.}(2008)\citenamefont {Ohashi},
  \citenamefont {Momoi}, \citenamefont {Tsunetsugu},\ and\ \citenamefont
  {Kawakami}}]{ohashi_finite_2008}%
  \BibitemOpen
  \bibfield  {author} {\bibinfo {author} {\bibfnamefont {T.}~\bibnamefont
  {Ohashi}}, \bibinfo {author} {\bibfnamefont {T.}~\bibnamefont {Momoi}},
  \bibinfo {author} {\bibfnamefont {H.}~\bibnamefont {Tsunetsugu}}, \ and\
  \bibinfo {author} {\bibfnamefont {N.}~\bibnamefont {Kawakami}},\ }\href
  {\doibase 10.1103/PhysRevLett.100.076402} {\bibfield  {journal} {\bibinfo
  {journal} {Phys. Rev. Lett.}\ }\textbf {\bibinfo {volume} {100}},\ \bibinfo
  {pages} {076402} (\bibinfo {year} {2008})}\BibitemShut {NoStop}%
\bibitem [{\citenamefont {Sahebsara}\ and\ \citenamefont
  {S{\'e}n{\'e}chal}(2006)}]{sahebsara_antiferromagnetism_2006}%
  \BibitemOpen
  \bibfield  {author} {\bibinfo {author} {\bibfnamefont {P.}~\bibnamefont
  {Sahebsara}}\ and\ \bibinfo {author} {\bibfnamefont {D.}~\bibnamefont
  {S{\'e}n{\'e}chal}},\ }\href {\doibase 10.1103/PhysRevLett.97.257004}
  {\bibfield  {journal} {\bibinfo  {journal} {Phys. Rev. Lett.}\ }\textbf
  {\bibinfo {volume} {97}},\ \bibinfo {pages} {257004} (\bibinfo {year}
  {2006})}\BibitemShut {NoStop}%
\bibitem [{\citenamefont {Laubach}\ \emph {et~al.}(2015)\citenamefont
  {Laubach}, \citenamefont {Thomale}, \citenamefont {Platt}, \citenamefont
  {Hanke},\ and\ \citenamefont {Li}}]{laubach_phase_2015}%
  \BibitemOpen
  \bibfield  {author} {\bibinfo {author} {\bibfnamefont {M.}~\bibnamefont
  {Laubach}}, \bibinfo {author} {\bibfnamefont {R.}~\bibnamefont {Thomale}},
  \bibinfo {author} {\bibfnamefont {C.}~\bibnamefont {Platt}}, \bibinfo
  {author} {\bibfnamefont {W.}~\bibnamefont {Hanke}}, \ and\ \bibinfo {author}
  {\bibfnamefont {G.}~\bibnamefont {Li}},\ }\href {\doibase
  10.1103/PhysRevB.91.245125} {\bibfield  {journal} {\bibinfo  {journal} {Phys.
  Rev. B}\ }\textbf {\bibinfo {volume} {91}},\ \bibinfo {pages} {245125}
  (\bibinfo {year} {2015})}\BibitemShut {NoStop}%
\bibitem [{\citenamefont {Yamada}(2014)}]{yamada_magnetic_2014}%
  \BibitemOpen
  \bibfield  {author} {\bibinfo {author} {\bibfnamefont {A.}~\bibnamefont
  {Yamada}},\ }\href {\doibase 10.1103/PhysRevB.89.195108} {\bibfield
  {journal} {\bibinfo  {journal} {Phys. Rev. B}\ }\textbf {\bibinfo {volume}
  {89}},\ \bibinfo {pages} {195108} (\bibinfo {year} {2014})}\BibitemShut
  {NoStop}%
\bibitem [{\citenamefont {Watanabe}\ \emph {et~al.}(2008)\citenamefont
  {Watanabe}, \citenamefont {Yokoyama}, \citenamefont {Tanaka},\ and\
  \citenamefont {Inoue}}]{watanabe_predominant_2008}%
  \BibitemOpen
  \bibfield  {author} {\bibinfo {author} {\bibfnamefont {T.}~\bibnamefont
  {Watanabe}}, \bibinfo {author} {\bibfnamefont {H.}~\bibnamefont {Yokoyama}},
  \bibinfo {author} {\bibfnamefont {Y.}~\bibnamefont {Tanaka}}, \ and\ \bibinfo
  {author} {\bibfnamefont {J.}~\bibnamefont {Inoue}},\ }\href {\doibase
  10.1103/PhysRevB.77.214505} {\bibfield  {journal} {\bibinfo  {journal} {Phys.
  Rev. B}\ }\textbf {\bibinfo {volume} {77}},\ \bibinfo {pages} {214505}
  (\bibinfo {year} {2008})}\BibitemShut {NoStop}%
\bibitem [{\citenamefont {Liebsch}\ \emph {et~al.}(2009)\citenamefont
  {Liebsch}, \citenamefont {Ishida},\ and\ \citenamefont
  {Merino}}]{liebsch_mott_2009}%
  \BibitemOpen
  \bibfield  {author} {\bibinfo {author} {\bibfnamefont {A.}~\bibnamefont
  {Liebsch}}, \bibinfo {author} {\bibfnamefont {H.}~\bibnamefont {Ishida}}, \
  and\ \bibinfo {author} {\bibfnamefont {J.}~\bibnamefont {Merino}},\ }\href
  {\doibase 10.1103/PhysRevB.79.195108} {\bibfield  {journal} {\bibinfo
  {journal} {Phys. Rev. B}\ }\textbf {\bibinfo {volume} {79}},\ \bibinfo
  {pages} {195108} (\bibinfo {year} {2009})}\BibitemShut {NoStop}%
\bibitem [{\citenamefont {Powell}\ and\ \citenamefont
  {McKenzie}(2007)}]{powell_symmetry_2007}%
  \BibitemOpen
  \bibfield  {author} {\bibinfo {author} {\bibfnamefont {B.~J.}\ \bibnamefont
  {Powell}}\ and\ \bibinfo {author} {\bibfnamefont {R.~H.}\ \bibnamefont
  {McKenzie}},\ }\href {\doibase 10.1103/PhysRevLett.98.027005} {\bibfield
  {journal} {\bibinfo  {journal} {Phys. Rev. Lett.}\ }\textbf {\bibinfo
  {volume} {98}},\ \bibinfo {pages} {027005} (\bibinfo {year}
  {2007})}\BibitemShut {NoStop}%
\bibitem [{\citenamefont {Metzner}\ and\ \citenamefont
  {Vollhardt}(1989)}]{metzner_correlated_1989}%
  \BibitemOpen
  \bibfield  {author} {\bibinfo {author} {\bibfnamefont {W.}~\bibnamefont
  {Metzner}}\ and\ \bibinfo {author} {\bibfnamefont {D.}~\bibnamefont
  {Vollhardt}},\ }\href {\doibase 10.1103/PhysRevLett.62.324} {\bibfield
  {journal} {\bibinfo  {journal} {Phys. Rev. Lett.}\ }\textbf {\bibinfo
  {volume} {62}},\ \bibinfo {pages} {324} (\bibinfo {year} {1989})}\BibitemShut
  {NoStop}%
\bibitem [{\citenamefont {Georges}\ and\ \citenamefont
  {Kotliar}(1992)}]{georges_hubbard_1992}%
  \BibitemOpen
  \bibfield  {author} {\bibinfo {author} {\bibfnamefont {A.}~\bibnamefont
  {Georges}}\ and\ \bibinfo {author} {\bibfnamefont {G.}~\bibnamefont
  {Kotliar}},\ }\href {\doibase 10.1103/PhysRevB.45.6479} {\bibfield  {journal}
  {\bibinfo  {journal} {Phys. Rev. B}\ }\textbf {\bibinfo {volume} {45}},\
  \bibinfo {pages} {6479} (\bibinfo {year} {1992})}\BibitemShut {NoStop}%
\bibitem [{\citenamefont {Georges}\ \emph {et~al.}(1996)\citenamefont
  {Georges}, \citenamefont {Kotliar}, \citenamefont {Krauth},\ and\
  \citenamefont {Rozenberg}}]{georges_dynamical_1996}%
  \BibitemOpen
  \bibfield  {author} {\bibinfo {author} {\bibfnamefont {A.}~\bibnamefont
  {Georges}}, \bibinfo {author} {\bibfnamefont {G.}~\bibnamefont {Kotliar}},
  \bibinfo {author} {\bibfnamefont {W.}~\bibnamefont {Krauth}}, \ and\ \bibinfo
  {author} {\bibfnamefont {M.~J.}\ \bibnamefont {Rozenberg}},\ }\href {\doibase
  10.1103/RevModPhys.68.13} {\bibfield  {journal} {\bibinfo  {journal} {Rev.
  Mod. Phys.}\ }\textbf {\bibinfo {volume} {68}},\ \bibinfo {pages} {13}
  (\bibinfo {year} {1996})}\BibitemShut {NoStop}%
\bibitem [{\citenamefont {Weiss}(1907)}]{weiss_hypothese_1907}%
  \BibitemOpen
  \bibfield  {author} {\bibinfo {author} {\bibfnamefont {P.}~\bibnamefont
  {Weiss}},\ }\href {\doibase 10.1051/jphystap:019070060066100} {\bibfield
  {journal} {\bibinfo  {journal} {J. Phys. Theor. Appl.}\ }\textbf {\bibinfo
  {volume} {6}},\ \bibinfo {pages} {661} (\bibinfo {year} {1907})}\BibitemShut
  {NoStop}%
\bibitem [{\citenamefont {Rokhsar}\ and\ \citenamefont
  {Kotliar}(1991)}]{rokhsar_gutzwiller_1991}%
  \BibitemOpen
  \bibfield  {author} {\bibinfo {author} {\bibfnamefont {D.~S.}\ \bibnamefont
  {Rokhsar}}\ and\ \bibinfo {author} {\bibfnamefont {B.~G.}\ \bibnamefont
  {Kotliar}},\ }\href {\doibase 10.1103/PhysRevB.44.10328} {\bibfield
  {journal} {\bibinfo  {journal} {Phys. Rev. B}\ }\textbf {\bibinfo {volume}
  {44}},\ \bibinfo {pages} {10328} (\bibinfo {year} {1991})}\BibitemShut
  {NoStop}%
\bibitem [{\citenamefont {Krauth}\ \emph {et~al.}(1992)\citenamefont {Krauth},
  \citenamefont {Caffarel},\ and\ \citenamefont
  {Bouchaud}}]{krauth_gutzwiller_1992}%
  \BibitemOpen
  \bibfield  {author} {\bibinfo {author} {\bibfnamefont {W.}~\bibnamefont
  {Krauth}}, \bibinfo {author} {\bibfnamefont {M.}~\bibnamefont {Caffarel}}, \
  and\ \bibinfo {author} {\bibfnamefont {J.-P.}\ \bibnamefont {Bouchaud}},\
  }\href {\doibase 10.1103/PhysRevB.45.3137} {\bibfield  {journal} {\bibinfo
  {journal} {Phys. Rev. B}\ }\textbf {\bibinfo {volume} {45}},\ \bibinfo
  {pages} {3137} (\bibinfo {year} {1992})}\BibitemShut {NoStop}%
\bibitem [{\citenamefont {L{\"u}hmann}(2013)}]{luhmann_cluster_2013}%
  \BibitemOpen
  \bibfield  {author} {\bibinfo {author} {\bibfnamefont {D.-S.}\ \bibnamefont
  {L{\"u}hmann}},\ }\href {\doibase 10.1103/PhysRevA.87.043619} {\bibfield
  {journal} {\bibinfo  {journal} {Phys. Rev. A}\ }\textbf {\bibinfo {volume}
  {87}},\ \bibinfo {pages} {043619} (\bibinfo {year} {2013})}\BibitemShut
  {NoStop}%
\bibitem [{\citenamefont {Yamamoto}\ \emph {et~al.}(2012)\citenamefont
  {Yamamoto}, \citenamefont {Danshita},\ and\ \citenamefont {S{\'a}~de
  Melo}}]{yamamoto_dipolar_2012}%
  \BibitemOpen
  \bibfield  {author} {\bibinfo {author} {\bibfnamefont {D.}~\bibnamefont
  {Yamamoto}}, \bibinfo {author} {\bibfnamefont {I.}~\bibnamefont {Danshita}},
  \ and\ \bibinfo {author} {\bibfnamefont {C.~A.~R.}\ \bibnamefont {S{\'a}~de
  Melo}},\ }\href {\doibase 10.1103/PhysRevA.85.021601} {\bibfield  {journal}
  {\bibinfo  {journal} {Phys. Rev. A}\ }\textbf {\bibinfo {volume} {85}},\
  \bibinfo {pages} {021601} (\bibinfo {year} {2012})}\BibitemShut {NoStop}%
\bibitem [{\citenamefont {Yamamoto}\ \emph {et~al.}(2014)\citenamefont
  {Yamamoto}, \citenamefont {Marmorini},\ and\ \citenamefont
  {Danshita}}]{yamamoto_quantum_2014}%
  \BibitemOpen
  \bibfield  {author} {\bibinfo {author} {\bibfnamefont {D.}~\bibnamefont
  {Yamamoto}}, \bibinfo {author} {\bibfnamefont {G.}~\bibnamefont {Marmorini}},
  \ and\ \bibinfo {author} {\bibfnamefont {I.}~\bibnamefont {Danshita}},\
  }\href {\doibase 10.1103/PhysRevLett.112.127203} {\bibfield  {journal}
  {\bibinfo  {journal} {Phys. Rev. Lett.}\ }\textbf {\bibinfo {volume} {112}},\
  \bibinfo {pages} {127203} (\bibinfo {year} {2014})}\BibitemShut {NoStop}%
\bibitem [{\citenamefont {Oguchi}(1955)}]{oguchi_theory_1955}%
  \BibitemOpen
  \bibfield  {author} {\bibinfo {author} {\bibfnamefont {T.}~\bibnamefont
  {Oguchi}},\ }\href {\doibase 10.1143/PTP.13.148} {\bibfield  {journal}
  {\bibinfo  {journal} {Prog. Theor. Phys.}\ }\textbf {\bibinfo {volume}
  {13}},\ \bibinfo {pages} {148} (\bibinfo {year} {1955})}\BibitemShut
  {NoStop}%
\bibitem [{\citenamefont {Hettler}\ \emph {et~al.}(1998)\citenamefont
  {Hettler}, \citenamefont {Tahvildar-Zadeh}, \citenamefont {Jarrell},
  \citenamefont {Pruschke},\ and\ \citenamefont
  {Krishnamurthy}}]{hettler_nonlocal_1998}%
  \BibitemOpen
  \bibfield  {author} {\bibinfo {author} {\bibfnamefont {M.~H.}\ \bibnamefont
  {Hettler}}, \bibinfo {author} {\bibfnamefont {A.~N.}\ \bibnamefont
  {Tahvildar-Zadeh}}, \bibinfo {author} {\bibfnamefont {M.}~\bibnamefont
  {Jarrell}}, \bibinfo {author} {\bibfnamefont {T.}~\bibnamefont {Pruschke}}, \
  and\ \bibinfo {author} {\bibfnamefont {H.~R.}\ \bibnamefont
  {Krishnamurthy}},\ }\href {\doibase 10.1103/PhysRevB.58.R7475} {\bibfield
  {journal} {\bibinfo  {journal} {Phys. Rev. B}\ }\textbf {\bibinfo {volume}
  {58}},\ \bibinfo {pages} {R7475} (\bibinfo {year} {1998})}\BibitemShut
  {NoStop}%
\bibitem [{\citenamefont {Lichtenstein}\ and\ \citenamefont
  {Katsnelson}(2000)}]{lichtenstein_antiferromagnetism_2000}%
  \BibitemOpen
  \bibfield  {author} {\bibinfo {author} {\bibfnamefont {A.~I.}\ \bibnamefont
  {Lichtenstein}}\ and\ \bibinfo {author} {\bibfnamefont {M.~I.}\ \bibnamefont
  {Katsnelson}},\ }\href {\doibase 10.1103/PhysRevB.62.R9283} {\bibfield
  {journal} {\bibinfo  {journal} {Phys. Rev. B}\ }\textbf {\bibinfo {volume}
  {62}},\ \bibinfo {pages} {R9283} (\bibinfo {year} {2000})}\BibitemShut
  {NoStop}%
\bibitem [{\citenamefont {Kotliar}\ \emph {et~al.}(2001)\citenamefont
  {Kotliar}, \citenamefont {Savrasov}, \citenamefont {P{\'a}lsson},\ and\
  \citenamefont {Biroli}}]{kotliar_cellular_2001}%
  \BibitemOpen
  \bibfield  {author} {\bibinfo {author} {\bibfnamefont {G.}~\bibnamefont
  {Kotliar}}, \bibinfo {author} {\bibfnamefont {S.~Y.}\ \bibnamefont
  {Savrasov}}, \bibinfo {author} {\bibfnamefont {G.}~\bibnamefont
  {P{\'a}lsson}}, \ and\ \bibinfo {author} {\bibfnamefont {G.}~\bibnamefont
  {Biroli}},\ }\href {\doibase 10.1103/PhysRevLett.87.186401} {\bibfield
  {journal} {\bibinfo  {journal} {Phys. Rev. Lett.}\ }\textbf {\bibinfo
  {volume} {87}},\ \bibinfo {pages} {186401} (\bibinfo {year}
  {2001})}\BibitemShut {NoStop}%
\bibitem [{\citenamefont {Anderson}(1952)}]{anderson_approximate_1952}%
  \BibitemOpen
  \bibfield  {author} {\bibinfo {author} {\bibfnamefont {P.~W.}\ \bibnamefont
  {Anderson}},\ }\href {\doibase 10.1103/PhysRev.86.694} {\bibfield  {journal}
  {\bibinfo  {journal} {Phys. Rev.}\ }\textbf {\bibinfo {volume} {86}},\
  \bibinfo {pages} {694} (\bibinfo {year} {1952})}\BibitemShut {NoStop}%
\bibitem [{\citenamefont {White}(1992)}]{white_density_1992}%
  \BibitemOpen
  \bibfield  {author} {\bibinfo {author} {\bibfnamefont {S.~R.}\ \bibnamefont
  {White}},\ }\href {\doibase 10.1103/PhysRevLett.69.2863} {\bibfield
  {journal} {\bibinfo  {journal} {Phys. Rev. Lett.}\ }\textbf {\bibinfo
  {volume} {69}},\ \bibinfo {pages} {2863} (\bibinfo {year}
  {1992})}\BibitemShut {NoStop}%
\bibitem [{\citenamefont {White}(1993)}]{white_density-matrix_1993}%
  \BibitemOpen
  \bibfield  {author} {\bibinfo {author} {\bibfnamefont {S.~R.}\ \bibnamefont
  {White}},\ }\href {\doibase 10.1103/PhysRevB.48.10345} {\bibfield  {journal}
  {\bibinfo  {journal} {Phys. Rev. B}\ }\textbf {\bibinfo {volume} {48}},\
  \bibinfo {pages} {10345} (\bibinfo {year} {1993})}\BibitemShut {NoStop}%
\bibitem [{\citenamefont
  {Schollw{\"o}ck}(2011)}]{schollwock_density-matrix_2011}%
  \BibitemOpen
  \bibfield  {author} {\bibinfo {author} {\bibfnamefont {U.}~\bibnamefont
  {Schollw{\"o}ck}},\ }\href {\doibase 10.1016/j.aop.2010.09.012} {\bibfield
  {journal} {\bibinfo  {journal} {Annals of Physics}\ }\bibinfo {series}
  {January 2011 {Special} {Issue}},\ \textbf {\bibinfo {volume} {326}},\
  \bibinfo {pages} {96} (\bibinfo {year} {2011})}\BibitemShut {NoStop}%
\bibitem [{\citenamefont {Wolf}\ \emph {et~al.}(2015)\citenamefont {Wolf},
  \citenamefont {Go}, \citenamefont {McCulloch}, \citenamefont {Millis},\ and\
  \citenamefont {Schollw{\"o}ck}}]{wolf_imaginary-time_2015}%
  \BibitemOpen
  \bibfield  {author} {\bibinfo {author} {\bibfnamefont {F.~A.}\ \bibnamefont
  {Wolf}}, \bibinfo {author} {\bibfnamefont {A.}~\bibnamefont {Go}}, \bibinfo
  {author} {\bibfnamefont {I.~P.}\ \bibnamefont {McCulloch}}, \bibinfo {author}
  {\bibfnamefont {A.~J.}\ \bibnamefont {Millis}}, \ and\ \bibinfo {author}
  {\bibfnamefont {U.}~\bibnamefont {Schollw{\"o}ck}},\ }\href {\doibase
  10.1103/PhysRevX.5.041032} {\bibfield  {journal} {\bibinfo  {journal} {Phys.
  Rev. X}\ }\textbf {\bibinfo {volume} {5}},\ \bibinfo {pages} {041032}
  (\bibinfo {year} {2015})}\BibitemShut {NoStop}%
\bibitem [{\citenamefont {Wolf}\ \emph {et~al.}(2014)\citenamefont {Wolf},
  \citenamefont {McCulloch},\ and\ \citenamefont
  {Schollw{\"o}ck}}]{wolf_solving_2014}%
  \BibitemOpen
  \bibfield  {author} {\bibinfo {author} {\bibfnamefont {F.~A.}\ \bibnamefont
  {Wolf}}, \bibinfo {author} {\bibfnamefont {I.~P.}\ \bibnamefont {McCulloch}},
  \ and\ \bibinfo {author} {\bibfnamefont {U.}~\bibnamefont {Schollw{\"o}ck}},\
  }\href {\doibase 10.1103/PhysRevB.90.235131} {\bibfield  {journal} {\bibinfo
  {journal} {Phys. Rev. B}\ }\textbf {\bibinfo {volume} {90}},\ \bibinfo
  {pages} {235131} (\bibinfo {year} {2014})}\BibitemShut {NoStop}%
\bibitem [{\citenamefont {Sarkar}\ and\ \citenamefont
  {Pereira}(1995)}]{sarkar_using_1995}%
  \BibitemOpen
  \bibfield  {author} {\bibinfo {author} {\bibfnamefont {T.}~\bibnamefont
  {Sarkar}}\ and\ \bibinfo {author} {\bibfnamefont {O.}~\bibnamefont
  {Pereira}},\ }\href {\doibase 10.1109/74.370583} {\bibfield  {journal}
  {\bibinfo  {journal} {IEEE Antennas and Propagation Magazine}\ }\textbf
  {\bibinfo {volume} {37}},\ \bibinfo {pages} {48} (\bibinfo {year}
  {1995})}\BibitemShut {NoStop}%
\bibitem [{\citenamefont {Fetter}\ and\ \citenamefont
  {Walecka}(2003)}]{fetter_quantum_2003}%
  \BibitemOpen
  \bibfield  {author} {\bibinfo {author} {\bibfnamefont {A.~L.}\ \bibnamefont
  {Fetter}}\ and\ \bibinfo {author} {\bibfnamefont {J.~D.}\ \bibnamefont
  {Walecka}},\ }\href@noop {} {\emph {\bibinfo {title} {Quantum {Theory} of
  {Many}-{Particle} {Systems}}}}\ (\bibinfo  {publisher} {Dover Publications},\
  \bibinfo {address} {Mineola, N.Y},\ \bibinfo {year} {2003})\BibitemShut
  {NoStop}%
\bibitem [{\citenamefont
  {S{\'e}n{\'e}chal}(2008)}]{senechal_introduction_2008}%
  \BibitemOpen
  \bibfield  {author} {\bibinfo {author} {\bibfnamefont {D.}~\bibnamefont
  {S{\'e}n{\'e}chal}},\ }\href {http://arxiv.org/abs/0806.2690} {\bibfield
  {journal} {\bibinfo  {journal} {arXiv:0806.2690 [cond-mat]}\ } (\bibinfo
  {year} {2008})},\ \bibinfo {note} {arXiv: 0806.2690}\BibitemShut {NoStop}%
\bibitem [{\citenamefont {Bernu}\ \emph {et~al.}(1994)\citenamefont {Bernu},
  \citenamefont {Lecheminant}, \citenamefont {Lhuillier},\ and\ \citenamefont
  {Pierre}}]{bernu_exact_1994}%
  \BibitemOpen
  \bibfield  {author} {\bibinfo {author} {\bibfnamefont {B.}~\bibnamefont
  {Bernu}}, \bibinfo {author} {\bibfnamefont {P.}~\bibnamefont {Lecheminant}},
  \bibinfo {author} {\bibfnamefont {C.}~\bibnamefont {Lhuillier}}, \ and\
  \bibinfo {author} {\bibfnamefont {L.}~\bibnamefont {Pierre}},\ }\href
  {\doibase 10.1103/PhysRevB.50.10048} {\bibfield  {journal} {\bibinfo
  {journal} {Phys. Rev. B}\ }\textbf {\bibinfo {volume} {50}},\ \bibinfo
  {pages} {10048} (\bibinfo {year} {1994})}\BibitemShut {NoStop}%
\bibitem [{\citenamefont {Capriotti}\ \emph {et~al.}(1999)\citenamefont
  {Capriotti}, \citenamefont {Trumper},\ and\ \citenamefont
  {Sorella}}]{capriotti_long-range_1999}%
  \BibitemOpen
  \bibfield  {author} {\bibinfo {author} {\bibfnamefont {L.}~\bibnamefont
  {Capriotti}}, \bibinfo {author} {\bibfnamefont {A.~E.}\ \bibnamefont
  {Trumper}}, \ and\ \bibinfo {author} {\bibfnamefont {S.}~\bibnamefont
  {Sorella}},\ }\href {\doibase 10.1103/PhysRevLett.82.3899} {\bibfield
  {journal} {\bibinfo  {journal} {Phys. Rev. Lett.}\ }\textbf {\bibinfo
  {volume} {82}},\ \bibinfo {pages} {3899} (\bibinfo {year}
  {1999})}\BibitemShut {NoStop}%
\bibitem [{\citenamefont {White}\ and\ \citenamefont
  {Chernyshev}(2007)}]{white_neorder_2007}%
  \BibitemOpen
  \bibfield  {author} {\bibinfo {author} {\bibfnamefont {S.~R.}\ \bibnamefont
  {White}}\ and\ \bibinfo {author} {\bibfnamefont {A.~L.}\ \bibnamefont
  {Chernyshev}},\ }\href {\doibase 10.1103/PhysRevLett.99.127004} {\bibfield
  {journal} {\bibinfo  {journal} {Phys. Rev. Lett.}\ }\textbf {\bibinfo
  {volume} {99}},\ \bibinfo {pages} {127004} (\bibinfo {year}
  {2007})}\BibitemShut {NoStop}%
\bibitem [{\citenamefont {Shirakawa}\ \emph {et~al.}(2016)\citenamefont
  {Shirakawa}, \citenamefont {Tohyama}, \citenamefont {Kokalj}, \citenamefont
  {Sota},\ and\ \citenamefont {Yunoki}}]{shirakawa_ground_2016}%
  \BibitemOpen
  \bibfield  {author} {\bibinfo {author} {\bibfnamefont {T.}~\bibnamefont
  {Shirakawa}}, \bibinfo {author} {\bibfnamefont {T.}~\bibnamefont {Tohyama}},
  \bibinfo {author} {\bibfnamefont {J.}~\bibnamefont {Kokalj}}, \bibinfo
  {author} {\bibfnamefont {S.}~\bibnamefont {Sota}}, \ and\ \bibinfo {author}
  {\bibfnamefont {S.}~\bibnamefont {Yunoki}},\ }\href
  {http://arxiv.org/abs/1606.06814} {\bibfield  {journal} {\bibinfo  {journal}
  {arXiv:1606.06814 [cond-mat]}\ } (\bibinfo {year} {2016})},\ \bibinfo {note}
  {arXiv: 1606.06814}\BibitemShut {NoStop}%
\bibitem [{\citenamefont {Sahebsara}\ and\ \citenamefont
  {S{\'e}n{\'e}chal}(2008)}]{sahebsara_hubbard_2008}%
  \BibitemOpen
  \bibfield  {author} {\bibinfo {author} {\bibfnamefont {P.}~\bibnamefont
  {Sahebsara}}\ and\ \bibinfo {author} {\bibfnamefont {D.}~\bibnamefont
  {S{\'e}n{\'e}chal}},\ }\href {\doibase 10.1103/PhysRevLett.100.136402}
  {\bibfield  {journal} {\bibinfo  {journal} {Phys. Rev. Lett.}\ }\textbf
  {\bibinfo {volume} {100}},\ \bibinfo {pages} {136402} (\bibinfo {year}
  {2008})}\BibitemShut {NoStop}%
\bibitem [{\citenamefont {Toschi}\ \emph {et~al.}(2007)\citenamefont {Toschi},
  \citenamefont {Katanin},\ and\ \citenamefont {Held}}]{toschi_dynamical_2007}%
  \BibitemOpen
  \bibfield  {author} {\bibinfo {author} {\bibfnamefont {A.}~\bibnamefont
  {Toschi}}, \bibinfo {author} {\bibfnamefont {A.~A.}\ \bibnamefont {Katanin}},
  \ and\ \bibinfo {author} {\bibfnamefont {K.}~\bibnamefont {Held}},\ }\href
  {\doibase 10.1103/PhysRevB.75.045118} {\bibfield  {journal} {\bibinfo
  {journal} {Phys. Rev. B}\ }\textbf {\bibinfo {volume} {75}},\ \bibinfo
  {pages} {045118} (\bibinfo {year} {2007})}\BibitemShut {NoStop}%
\bibitem [{\citenamefont {Rubtsov}\ \emph {et~al.}(2008)\citenamefont
  {Rubtsov}, \citenamefont {Katsnelson},\ and\ \citenamefont
  {Lichtenstein}}]{rubtsov_dual_2008}%
  \BibitemOpen
  \bibfield  {author} {\bibinfo {author} {\bibfnamefont {A.~N.}\ \bibnamefont
  {Rubtsov}}, \bibinfo {author} {\bibfnamefont {M.~I.}\ \bibnamefont
  {Katsnelson}}, \ and\ \bibinfo {author} {\bibfnamefont {A.~I.}\ \bibnamefont
  {Lichtenstein}},\ }\href {\doibase 10.1103/PhysRevB.77.033101} {\bibfield
  {journal} {\bibinfo  {journal} {Phys. Rev. B}\ }\textbf {\bibinfo {volume}
  {77}},\ \bibinfo {pages} {033101} (\bibinfo {year} {2008})}\BibitemShut
  {NoStop}%
\bibitem [{\citenamefont {Go}\ and\ \citenamefont
  {Millis}(2015)}]{go_spatial_2015}%
  \BibitemOpen
  \bibfield  {author} {\bibinfo {author} {\bibfnamefont {A.}~\bibnamefont
  {Go}}\ and\ \bibinfo {author} {\bibfnamefont {A.~J.}\ \bibnamefont
  {Millis}},\ }\href {\doibase 10.1103/PhysRevLett.114.016402} {\bibfield
  {journal} {\bibinfo  {journal} {Phys. Rev. Lett.}\ }\textbf {\bibinfo
  {volume} {114}},\ \bibinfo {pages} {016402} (\bibinfo {year}
  {2015})}\BibitemShut {NoStop}%
\bibitem [{\citenamefont {Gustavsen}\ and\ \citenamefont
  {Semlyen}(1999)}]{gustavsen_rational_1999}%
  \BibitemOpen
  \bibfield  {author} {\bibinfo {author} {\bibfnamefont {B.}~\bibnamefont
  {Gustavsen}}\ and\ \bibinfo {author} {\bibfnamefont {A.}~\bibnamefont
  {Semlyen}},\ }\href {\doibase 10.1109/61.772353} {\bibfield  {journal}
  {\bibinfo  {journal} {IEEE Transactions on Power Delivery}\ }\textbf
  {\bibinfo {volume} {14}},\ \bibinfo {pages} {1052} (\bibinfo {year}
  {1999})}\BibitemShut {NoStop}%
\bibitem [{\citenamefont {Gustavsen}\ and\ \citenamefont
  {Heitz}(2008)}]{gustavsen_modal_2008}%
  \BibitemOpen
  \bibfield  {author} {\bibinfo {author} {\bibfnamefont {B.}~\bibnamefont
  {Gustavsen}}\ and\ \bibinfo {author} {\bibfnamefont {C.}~\bibnamefont
  {Heitz}},\ }\href {\doibase 10.1109/TADVP.2008.927810} {\bibfield  {journal}
  {\bibinfo  {journal} {IEEE Transactions on Advanced Packaging}\ }\textbf
  {\bibinfo {volume} {31}},\ \bibinfo {pages} {664} (\bibinfo {year}
  {2008})}\BibitemShut {NoStop}%
\end{thebibliography}%
\end{document}